\documentclass[aps,pra,reprint,10pt,a4paper,superscriptaddress,nofootinbib]{revtex4-1}
\usepackage[utf8]{inputenc}
\usepackage[sc,osf]{mathpazo}
\usepackage{amsmath}
\usepackage[T1]{fontenc}
\usepackage{latexsym}
\usepackage{amssymb}
\usepackage{color}
\usepackage{graphics,epstopdf}
\usepackage{soul}
\usepackage[demo]{graphicx}
\usepackage{capt-of}
\usepackage{lipsum}
\usepackage{adjustbox}
\usepackage[normalem]{ulem}
\usepackage[table,xcdraw]{xcolor}
\usepackage{braket}
\usepackage{amsthm}
\usepackage{physics}
\usepackage{ragged2e}
\usepackage{mathtools}

\usepackage{comment}
\usepackage{ragged2e}
\usepackage{braket}
\usepackage[colorlinks=true,allcolors=blue]{hyperref}
\usepackage{natbib}

\usepackage{orcidlink}
\usepackage[justification=justified]{caption}
\usepackage[font=small,labelfont=bf,justification=justified,singlelinecheck=false]{caption}

\definecolor{indiagreen}{rgb}{0.07, 0.53, 0.03}
\definecolor{teal}{rgb}{0.0, 0.53, 0.53}

\begin{document}

\title{ 
Characterizing pairwise swapping capabilities of dense coding channels}


\author{Abhishek Muhuri \orcidlink{0000-0002-6763-2796}}
\email{abhishek.muhuri@research.iiit.ac.in}
\affiliation{Harish-Chandra Research Institute, Chhatnag Road, Jhunsi, Prayagraj - 211019, India}
\affiliation{Homi Bhabha National Institute, Training School Complex, Anushakti Nagar, Mumbai 400 094, India}
\affiliation{Center for Quantum Science and Technology, International Institute of Information
Technology Hyderabad, Gachibowli, Hyderabad, Telangana 500032, India}

\author{Nirman Ganguly \orcidlink{0009-0002-3232-6227}}
\email{nirmanganguly@hyderabad.bits-pilani.ac.in, nirmanganguly@gmail.com}
\affiliation{Department of Mathematics, Birla Institute of Technology and Science, Pilani,
Hyderabad Campus, Jawahar Nagar, Kapra Mandal, Medchal District, Telangana 500078, India}

\author{Aditi Sen(De) \orcidlink{0000-0003-1693-0440}}
\email{aditi@hri.res.in}
\affiliation{Harish-Chandra Research Institute, Chhatnag Road, Jhunsi, Prayagraj - 211019, India}
\affiliation{Homi Bhabha National Institute, Training School Complex, Anushakti Nagar, Mumbai 400 094, India}

\author{Indranil Chakrabarty \orcidlink{0009-0001-0415-0431}}
\email{indranil.chakrabarty@iiit.ac.in}
\affiliation{Centre for Quantum Science and Technology and Center for Security, Theory \& Algorithmic Research,  International Institute of Information
Technology Hyderabad, Gachibowli, Hyderabad, Telangana 500032, India}


\begin{abstract}

We introduce a novel multipartite entanglement-assisted classical communication task, referred to as  {\it dense coding swapping}, in which legitimate parties collaboratively swap the dense codeability from one communication channel to another through suitable joint unitary operations. Due to the dense coding (DC) exclusion principle,  the scheme enhances the dense codeability of a target pair while simultaneously reducing it for a non-target branch in the network. This swapping capability has broader implications, as it may be viewed as a form of process swapping, distinct from resource swapping, while also providing a prevention measure when one of the receivers is compromised.  We derive necessary and sufficient conditions, expressed in terms of the Schmidt coefficients, for three-qubit pure states to support DC swapping, while we obtain a sufficient criterion for mixed states using their Bloch correlation parameters. Furthermore, we identify the optimal two-qubit unitary operators capable of realizing the swapping of dense codeability between communication channels. We further examine the tolerance of these
eligible states against both colored and white noise, demonstrating the resilience of the proposed
task under environmental perturbations. We also show that multipartite states supporting DC swapping require only a small amount of genuine multipartite entanglement and that this requirement decreases with increasing system size.

 
\end{abstract}
\maketitle

\section{Introduction } 

Quantum conditional entropy is one of the most distinctive quantities in quantum information theory, as it can attain negative values, in sharp contrast to the classical realm. In quantum information science, this negativity~\cite{cerf1997negative,cerf1999quantum,friis2017geometry} has emerged as a fundamental resource, enabling several uniquely quantum information-processing tasks, including superdense coding~\cite{bennett1992communication,bruss93sen}, quantum state merging~\cite{horodecki2007quantum,horodecki2005partial}, one-way entanglement distillation~\cite{devetak2005distillation}, and the characterization of quantum channel capacities through coherent information~\cite{gyongyosi2012properties}. 
Beyond quantum communication, its impact extends to diverse areas of physics. It plays a crucial role in strengthening entropic uncertainty relations in the presence of quantum memory~\cite{berta2010uncertainty}, determining the thermodynamic work cost of quantum systems~\cite{rio2011thermodynamic}, and even modifying the Bekenstein--Hawking area law for Schwarzschild black holes~\cite{azuma2018black,azuma2020second}. These diverse applications highlight its fundamental importance across quantum information theory, quantum thermodynamics, and gravitational physics.\\
\noindent The significance of states with negative quantum conditional entropy led to the development of a resource-theoretic framework for their characterization and manipulation. A first step in this direction was the construction of Hermitian witnesses for such states, made possible by the convexity and compactness of the set of states with non-negative conditional von Neumann entropy (CVENN)~\cite{vempati2021witnessing}. Analogous to the resource theory of entanglement,
where separable states are free states and local operations and classical communication (LOCC)~\cite{vedral1997quantifying,horodecki2009quantum} are the free operations, the CVENN states form the free states in the resource theory of conditional von Neumann entropy, and the corresponding free operations are also identified~\cite{vempati2022unital}. This framework further admits an absolute class of states whose conditional entropy remains non-negative under arbitrary global unitary transformations~\cite{patro2017non}, referred to as the \emph{absolute conditional von Neumann entropy non-negative} (ACVENN) class, closely paralleling the notion of absolutely separable states~\cite{zyczkowski1998volume}. Moreover,
similar to the spirit of entanglement
breaking~\cite{horodecki2003entanglement,moravvcikova2010entanglement} and annihilating channels~\cite{moravvcikova2010entanglement}, there are channels that either break the negativity of the conditional entropies~\cite{srinidhi2024quantum,muhuri2023information},  or annihilate it ~\cite{srinidhi2024quantum}.\\ 
Superdense coding is a quantum communication protocol in which classical information encoded in a quantum state is transmitted between two distant parties, Alice and Bob, using a shared entangled quantum state, thereby achieving a communication capacity that surpasses the classical limit~\cite{bennett1992communication,srivastava2019one,das2015distributed,roy2018deterministic,bruss2006dense,nepal2013maximally,das2014multipartite,das2015distributed,samanta2024continuous,hao2001controlled,zhang2002controlled,jing2003experimental,fu2006controlled,oh2017minimal}. A bipartite quantum state provides such a quantum advantage if its conditional von Neumann entropy is negative~\cite{Hausladen_PRA_1996,Bowen_PRA_2001,Horodecki_arxiv_2001,prabhu2012exclusion}. Consequently, not every entangled state is useful for dense coding (DC), since entangled states belonging to the CVENN class possess non-negative conditional entropy, thereby offering no quantum benefit~\cite{patro2017non}. Over the years, several variants of the protocol, including multipartite~\cite{bruss2006dense,bruss93sen}, port-based~\cite{Nepal_PRA_2013,srivastava2019one}, and probabilistic dense coding~\cite{pati2005probabilistic}, have been developed, and their robustness against different noise models has been extensively investigated~\cite{Shadman_NJP_2010,Shadman2011,Shadman_PRA_2012,Shadman2013, das2014multipartite}. In multipartite settings, DC is further constrained by the exclusion principle, which states that no two bipartite reduced states of an \(N\)-party system sharing a common party can simultaneously exhibit a quantum advantage~\cite{prabhu2012exclusion}. Equivalently, the corresponding conditional entropies cannot both be negative at the same time. In this work, we demonstrate that this restriction is not immutable: by employing suitable two-qubit quantum gates, the dense codeability can be swapped from one pair of parties to another within a multipartite quantum state.\\
\noindent We propose a multipartite communication scenario in which Alice, acting as the controller, initially shares an entangled state with several parties such that, owing to the DC exclusion principle, only one communication channel possesses a quantum advantage at a given time. We address the following question: "{\it Can Alice, by performing a suitable unitary operation jointly with a target party, swap the DC advantage from the initially active channel to another desired channel?}" Since the exclusion principle forbids two channels sharing a common party from simultaneously exhibiting a quantum advantage, activating a previously inactive channel necessarily deactivates the initially active one (see Fig.~\ref{fig:ResourceStates} for the schematic illustration ). We answer this question affirmatively and refer to this phenomenon as {\it dense coding swapping}. This swapping of the dense coding channel becomes important when one of the receivers is compromised, providing an additional layer of operational robustness. This can be considered to be a first step toward the concept of process swapping, in contrast to resource-swapping protocols such as entanglement swapping~\cite{Zukowski_PRL_1993, Bose_PRA_1998}. \\
\noindent We prove necessary and sufficient conditions for three-qubit pure states, expressed in terms of the Schmidt coefficients, and we obtain the sufficient condition for the mixed three-party state via the correlation matrix,  under which DC swapping can be achieved using a two-qubit unitary operation. We illustrate this result by considering the three-qubit generalized W (gW) states.  Our numerical simulations reveal that three-, four-, and five-qubit Dicke states with one excitation also support such a swapping task. We further show that these resource states possess relatively small genuine multipartite entanglement, quantified by the generalized geometric measure (GGM)~\cite{Aditiggm_PRA_2010}, and perform a statistical analysis of their activation properties. In addition, we investigate the tolerance of the protocol to noise by determining the minimum noise strength at which tripartite resource states cease to support this activation. Furthermore, we derive explicit conditions on the two-qubit unitary operators required to activate the DC channel between Alice and the desired target party for both pure and noisy generalized three-qubit (W) states.\\
\noindent Our work is organized as follows: In Sec.  \ref{sec:switching}, we briefly discuss the idea of swapping the ability of the dense codeable channels in a multipartite setting. In Sec.  \ref{sec:charac}, we identify criteria for three-qubit pure and mixed states that are useful for DC swapping. Sec. \ref{sec:morequbitandGGM}
is devoted to characterizing three-and more-qubitstates for DC swapping and also establishing a connection between the multipartite entanglement measure and DC swapping. 
In Sec. \ref{sec:robustness}, we probe the tolerance of useful states by mixing them with white and orthogonal colored noise. Sec. \ref{sec:unitary} discusses the structure of unitaries which facilitates activation of dense coding. The concluding remarks are in  Sec. \ref{sec:conclusion}.

\section{Swapping Capabilities of Dense Coding Channels} 
\label{sec:switching}

\begin{figure}[t]
    \includegraphics[width=1.0\linewidth]{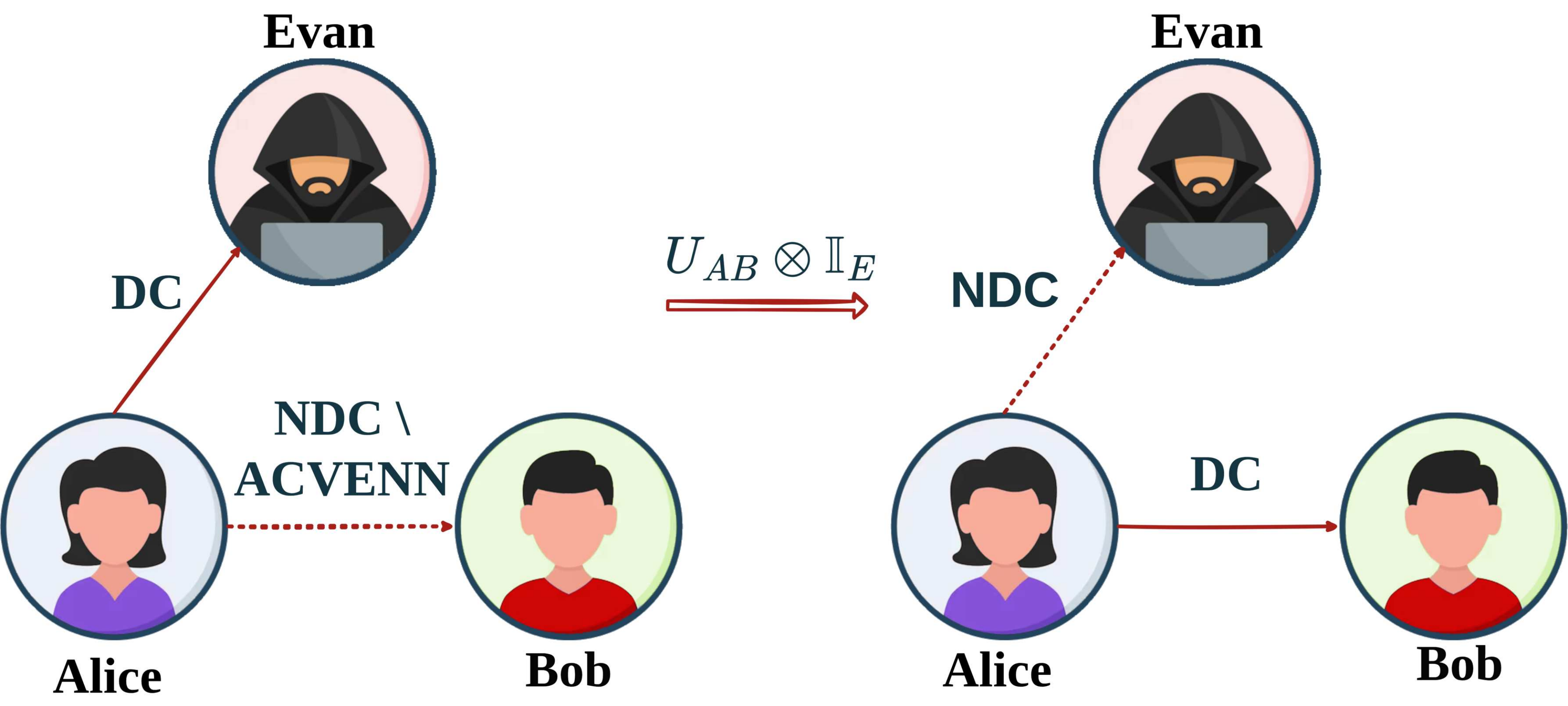}
    \caption{\justifying{{\bf Schematic diagram of the dense coding swapping in a tripartite setting.}
A tripartite state is shared among Alice (transmitter), Bob (desired receiver), and Evan (compromised party). Initially, Alice shares a dense codeable state with Evan, which precludes dense codeability between Alice and Bob due to the DC exclusion principle. If Evan gets compromised, Alice aims to stop granting quantum advantages to Evan and give the benefits to Bob by activating dense codeability in the channel $\rho_{AB}$ without invoking Evan. The swapping occurs when a suitable global unitary operation $U^{AB}$ transforms the reduced state shared between Alice and Bob into a dense codeable one, thereby enabling dense coding between them.}}
    \label{fig:ResourceStates}
\end{figure}

We propose a framework for dense coding swapping, namely the swapping of dense codeability between different pairs of channels in a multipartite state using two-qubit quantum gates applied on the desired pair. Before describing it, we first quantify the performance of the dense coding task.\\
\noindent {\it Dense coding capacity.}  The superdense coding capacity of a shared quantum state $\rho_{AB}$ with \(A\) and \(B\) being the sender and the receiver, respectively, reads as \cite{bruss93sen,bruss2006dense} $\mathcal{C}(\rho_{AB})= \max[\log_2 d_A, \log_2 d_A +S(\rho_{B})-S(\rho_{AB})]$, where \(S(\sigma) = -\Tr(\sigma \log_2 \sigma)\) is the Von Neumann entropy of \(\sigma\), \(d_A\) is the dimension of the sender's side, and \(\rho_B\) is the local density matrix of \(\rho_{AB}\).  Note that the classical limit of the DC protocol is  $\log_2 d_A$. The state is said to be dense codeable when  $S(\rho_{B})-S(\rho_{AB}) >0$. As a result, when the conditional entropy, \(S(A|B) = S(\rho_{AB}) - S(\rho_{B})\) is negative, the DC capacity is more than $\log_2 d_A$, which is referred to as the "quantum advantage" in DC. Otherwise, the state can be called non-dense codeable (NDC).\\ 
{\it Dense coding swapping.} Let us consider an N-qubit state, \(\rho_{AB_1\ldots B_{N-2}E}\), where one of the pairs, say \(\rho_{AE}\), has quantum advantage in DC while the rest of the pair (say $\rho_{AB_1}$ ) is NDC. In this protocol, we also assume that \(A\) is the boss and the sender, thereby controlling the dense coding protocol.   Once \(A\) confirms that \(E\) is dishonest, \(A\) can stop providing quantum benefits in classical information transmission to \(E\) and wishes to provide this to any of the \(B_i\)s, say \(B_1\) with the cooperation of \(B_1\) and without informing \(E\).   Further, we know  from the dense coding exclusion principle that \cite{prabhu2012exclusion}:
 {\it Given an arbitrary \(N\)-party quantum state $\rho_{AB_1\ldots B_{N-2}E}$, among the \(N-1\) pairs sharing a common party, at most one pair can exhibit a quantum advantage in the dense coding protocol.}
At some point, if $A$ finds the party $E$ to be compromised,  $A$ needs to stop granting quantum advantage in sending classical information to $E$, by making the channel $AE$ NDC so that it can never act as a resource, and at the same time, \(A\) wants to activate the channel shared between \(A\) and a target party \(B_1\). Moreover, this has to be achieved by $A$ and $B_1$ without invoking $E$. In particular, by applying a global unitary $U^{AB_1}$ on $A$ and $B_1$, $\rho_{AB_1}$ can be made dense codeable, which can transform $\rho_{AE}$ NDC  as envisioned by the exclusion principle. 
In other words, our goal is to  {\it swap} the quantum advantage  from \(\rho_{AE}\) to \(\rho_{AB_1}\) after the application of a suitable unitary on the composite system $AB_1$, i.e., we want to characterize the set of \(N\)-party states, such that
\begin{eqnarray}
   && \{\rho_{AB_1\ldots B_{N-2}E}| \mathcal{C}(\rho_{AE})>\log_2 d_A, \mathcal{C}(\rho_{AB_i})= \log_2 d_A \forall i\} \nonumber\\
  &&  \xrightarrow{U^{AB_1}} \nonumber \\
  &&\{\rho'_{AB_1\ldots B_{N-2}E}| \mathcal{C}(\rho'_{AB_1})>\log_2 d_A,  \mathcal{C}(\rho'_{AB_i/E})\nonumber\\&&= \log_2 d_A (i\neq t) \}  
  \label{eq:switch}
\end{eqnarray}
We refer to this task of swapping the dense codeability on $\rho_{AB}$ without involving $E$ through a unitary application as \textit{dense coding swapping} (for schematics, see Fig. \ref{fig:ResourceStates} for a three-party shared state). 

\noindent At this point, it becomes imperative to classify such \(N\)-qubit states, for which swapping is possible. This entails the identification of a novel kind of states which may be important in the context of secure information processing and thereby assumes significance in the realm of communication in multipartite quantum networks. The useful states are characterized by three specific conditions:
\begin{enumerate}
    \item $\rho_{AB_i} = \Tr_{\overline{AB_i}}[\rho_{AB_1\ldots B_{N-2}E}] \, \forall i$ is non-dense codeable, where \(\overline{AB_i}\) denotes  the partial trace being taken over all the parties except \(A\) and \(B_i\),  
    \item $\rho_{AB_1}$ $\notin$ \textit{ACVENN} \cite{patro2017non}, where $t$ indexes the target receiver and
    \item $\rho_{AE} = \Tr_{\overline{AE}}[\rho_{AB_1\ldots B_{N-2}E}]$ is dense codeable (\textit{DC}).
\end{enumerate}
Condition 2 is important, as belonging to  \textit{ACVENN} entails that there exists no unitary that can act on the composite system $AB_i$ and make it dense codeable.  The states that do not follow the above conditions 
cannot be used for dense coding swapping, and the set of these states is designated by $\mathcal{F}$. For illustration, we classify the three-qubit pure and mixed states according to their benefits of dense coding swapping in the next section. 


\noindent Another objective here is to characterize the unitary operators, \(U^{AB_1}\), which can swap the dense codeability from \(\rho_{AE}\) to \(\rho_{AB_1}\). To do so, we can use the parameterization of \(SU(4)\) \cite{Zhang_PRA_2003} and derive explicit conditions on parameters that are useful for dense coding swapping, especially for three-qubits (see Sec. \ref{sec:unitary}).


\section{Characterizing three-qubit resources for dense coding swapping} 
\label{sec:charac}

This section is devoted to deriving the conditions on the state parameters of three-qubit states under which such a transformation given in Eq. (\ref{eq:switch}) is possible. We begin with pure three-qubit states and characterize their usefulness for DC swapping in terms of the Schmidt coefficients. We then show that the mixed states can also be classified according to their DC swapping capability with the help of the correlation matrix. 


\subsection{Three-qubit pure states as a resource for DC swapping}
\label{subsec:pure}

Let us begin with the most general expression of three-qubit pure states~\cite{acin2001classification},
\begin{eqnarray}
\ket{\psi}_{ABE} =&& \lambda_0 \ket{000}+\lambda_1 e^{i\phi} \ket{100}+{}\nonumber\\&&\lambda_2 \ket{101}+\lambda_3 \ket{110}+\lambda_4 \ket{111},
\label{Equation1}
\end{eqnarray}
where \(\lambda_i\) (\(i \in \{0,1, \ldots 4\}\)) are real numbers referred to as the Schmidt coefficients and  \(0\leq \phi \leq 2 \pi\) represents the phase. Given that the above three-qubit state is shared by three parties, there can be three possible cases: (1) When $AB$ and $AE$ are both non-dense codeable; (2) when $AB$ is NDC but $AE$ is DC, and (3) when $AB$ is DC and $AE$ is NDC. 
Let us now concentrate on the first case and provide the conditions on \(\lambda_i\)s of \(|\psi\rangle_{ABE}\). In particular, we arrive at the necessary and sufficient conditions for non-dense codeabillity of \(AB\) and \(AE\).\\ 



\noindent \textbf{Theorem 1:} \textit{When the bipartition \(A:BE\) of a tripartite pure state is not separable, permutational symmetry between the parties \(B\) and \(E\) is necessary and sufficient to guarantee non-dense codeability of both the reduced states $\rho_{AB}=Tr_E(\ketbra{\psi}{\psi}_{ABE})$ and $\rho_{AE}=Tr_B(\ketbra{\psi}{\psi}_{ABE})$.}

\begin{proof}
      For the necessary part, we need to prove that if both $\rho_{AB}$ and $\rho_{AE}$ are not dense codeable,  the only possibility is $\ket{\psi}_{ABE}$ to be symmetric under the exchange of \(B\) and \(E\). We prove it by contradiction between the following statements:\\
      
\noindent \textbf{S$\mathbf{1.}$} When both the reduced states are not dense codeable,  both $S(Tr_{AB}(\ketbra{\psi}{\psi}_{ABE}))-S(Tr_{B}(\ketbra{\psi}{\psi}_{ABE})) = S(\rho_E) - S(\rho_{AE})\leq 0 $ and 
 $S(\rho_B) - S(\rho_{AB})\leq 0$.\\
\textbf{S$\mathbf{2.}$} $\ket{\psi}_{ABE}$ is permutationally asymmetric between parties $B$ and  $E$ , i.e., $\lambda_2 \neq \lambda_3$. Further,  \(\ket{\psi}_{A:BE}\) is non-separable,  when $\lambda_0 \neq 0$.

\noindent Assuming \textbf{S$\mathbf{1}$} is true when \textbf{S$\mathbf{2}$} is true, we find out the eigenvalues of the reduced states $\rho_{AB}$ and $\rho_{AE}$:\\
\begin{eqnarray}
\rho_{AB}: &&\Biggl\{0,0,\frac{1}{2} \pm \frac{1}{2}\Big[1+4\Bigl(\lambda_2^4+2\lambda_1\lambda_2\lambda_3\lambda_4\cos{\phi}+\nonumber\\
&&{+\lambda_4^2(-1+\lambda_3^2+\lambda_4^2)+\lambda_2^2(-1+\lambda_1^2+2\lambda_4^2)\Bigr)\Big]^{\frac{1}{2}}}\Biggr\}\nonumber,
\end{eqnarray}
and 
\begin{eqnarray}
\rho_{AE}: &&\Biggl\{0,0, \frac{1}{2} \pm \frac{1}{2}
    \Big[1 + 4\Bigl(\lambda_3^4 + 2\lambda_1\lambda_2\lambda_3\lambda_4 \cos{\phi}\nonumber\\
    && {+\lambda_4^2(-1 + \lambda_2^2 + \lambda_4^2)+ \lambda_3^2(-1 + \lambda_1^2 + 2\lambda_4^2)
    \Bigr)}\Big]^{\frac{1}{2}} \Biggr\}.\nonumber \,\,\\,
\end{eqnarray}
Since \(\ket{\psi}_{ABE}\) is pure, \(S(\rho_{AE}) = S(\rho_B)\) and \(S(\rho_{AB}) = S(\rho_E)\), and the spectra of the reduced states $\rho_B$ and $\rho_E$ are given by
\begin{eqnarray}
\label{eq:rhoB}
    \rho_B:  &&\Biggl\{ \frac{1}{2} \pm \frac{1}{2}
    \Big[1 + 4\Bigl(\lambda_3^4 + 2\lambda_1\lambda_2\lambda_3\lambda_4 \cos{\phi}\nonumber\\
     &&{+\lambda_4^2(-1 + \lambda_2^2 + \lambda_4^2)+ \lambda_3^2(-1 + \lambda_1^2 + 2\lambda_4^2)
    \Bigr)\Big]^{\frac{1}{2}}} \Biggr\},\nonumber \,\,\\
\end{eqnarray}
and 
\begin{eqnarray}
\label{eq:rhoE}
    \rho_{E}:  &&\Biggl\{\frac{1}{2} \pm \frac{1}{2}\Big[1+4\Bigl(\lambda_2^4+2\lambda_1\lambda_2\lambda_3\lambda_4\cos{\phi}+\nonumber\\
&&{+\lambda_4^2(-1+\lambda_3^2+\lambda_4^2)+\lambda_2^2(-1+\lambda_1^2+2\lambda_4^2)\Bigr)\Big]^{\frac{1}{2}}}\Biggr\}\nonumber.\\
\end{eqnarray}

It follows that $S(A|B)=-S(A|E)$, thus implying that both of the reduced density matrices can be simultaneously non-dense codeable unless both the conditional entropies vanish. This happens only when 
\begin{eqnarray}
    &&\lambda_3^4 +\lambda_4^2(-1 + \lambda_2^2 + \lambda_4^2)+ \lambda_3^2(-1 + \lambda_1^2 + 2\lambda_4^2)\nonumber \\
    &&=\lambda_2^4+\lambda_4^2(-1+\lambda_3^2+\lambda_4^2)+\lambda_2^2(-1+\lambda_1^2+2\lambda_4^2)\nonumber\\
    &\implies &\lambda_3^4 +\lambda_4^2\lambda_2^2+ \lambda_3^2(-1 + \lambda_1^2 + 2\lambda_4^2)\nonumber\\
    &&=\lambda_2^4+\lambda_4^2\lambda_3^2+\lambda_2^2(-1+\lambda_1^2+2\lambda_4^2)\nonumber\\
    &\implies& (\lambda_2^2-\lambda_3^2)\Big[\lambda_2^2+\lambda_3^2+\lambda_4^2+\lambda_1^2-1\Big] = 0\nonumber\\
    &\implies& \lambda_0 = 0 \text{ or } \lambda_2 = \lambda_3. 
\end{eqnarray}
The last line clearly contradicts the assumption \textbf{S$\mathbf{2}$} which demands that the state $\ket{\psi}_{ABE}$ needs to be permutationally symmetric between the parties $B$ and $E$. 

\noindent To prove the sufficient part, if $\lambda_2=\lambda_3$, the state \(|\psi\rangle_{ABE}\), is symmetric between the parties $B$ and $E$, and hence under this condition, the reduced states $\rho_{AE}$ and $\rho_{AB}$ are the same thus the DC capacities are equal. But due to the exclusion principle of dense coding, both pairs cannot be simultaneously dense codeable of a three-party state, i.e., under this condition, \(\rho_{AB}\) and \(\rho_{AE}\) are NDC. Hence proved.

\end{proof} 

\noindent Let us now move to the second and third situations, where one of the pairs, say \(\rho_{AE}\) of \(\ket{\psi}_{ABE}\) is DC and the other pair \(\rho_{AB}\) is NDC. Our goal is to find the necessary and sufficient condition based on the Schmidt coefficients of  \(\ket{\psi}_{ABE}\) for which we can deactivate Alice ($A$) to Evan's ($E$) channel and activate the Alice to Bob ($B$) channel. In order to determine this, we first find the set of states \(\ket{\psi}_{ABE}\) for which $\rho_{AE}$ is  DC and $\rho_{AB}$ is NDC and 
then we identify \(\ket{\psi}_{ABE}\)s for which $\rho_{AB} \notin ACVENN$. This is because if $\rho_{AB} \in ACVENN$, there exists no unitary which can make $\rho_{AB}$ useful for dense coding \cite{patro2017non}. Note that although we discuss the case (2), the condition for (3) can be derived in the similar line. Let us first derive when \(\mathcal{C}(\rho_{AB}) = \log_2 d_A\) and \(\mathcal{C}(\rho_{AB}) > \log_2 d_A\).\\ 



\noindent \textbf{Lemma $1.$} \textit{Any tripartite pure states written in a Schmidt decomposition for which $\rho_{AE}$ is dense codeable and $\rho_{AB}$ is non-dense codeable must satisfy the condition  $|\lambda_2|>|\lambda_3|$ provided $\lambda_0 \neq 0$ }. 

\begin{proof}    
For pure states, we have $S(\rho_{AB}) = S(\rho_E)$ and $S(\rho_{AE}) = S(\rho_B)$. Thus,  $S(A|B) + S(A|E) = 0$. Thus, it is imperative that if one is strictly positive,  the other would certainly be strictly negative. One might notice that the positivity of the conditional von-Neumann entropy of a bipartite state implies that the quantum advantage obtained during dense coding-based communication is vanishing. Thus, if the state exhibits positive or zero conditional von-Neumann entropy, the state is not dense codeable and vice versa. The previous relation between the conditional von-Neumann entropy between the parties thus holds a crucial point. $S(A|E)$ should be negative, correspondingly $S(A|B)$ is positive for the tripartite state $\ket{\psi}_{ABE}$ to design in a way that $\rho_{AE}$ is dense codeable, while it $\rho_{AB}$ is not. Not only that, $\rho_{AE}$ should also have strictly negative conditional von-Neumann entropy, while $\rho_{AB}$ should have strictly positive conditional von-Neumann entropy.  This is because, if one vanishes, the other would too, resulting in the non-dense codeability of both the reduced states. This implies we  need a state that has $S(A|B)>0$, which is basically $S(\rho_B)<S(\rho_{AB})$, and from earlier, we obtain $S(\rho_{AB}) = S(\rho_E)$, resulting into $S(\rho_B)<S(\rho_E)$. As both the reduced density matrices $\rho_B$ and $\rho_E$ are qubits, the entropy is binary. So, without loss of generality, suppose, $\rho_B$ has eigenvalues $\{\frac{1}{2}\pm e_b\}$ and $\rho_E$ has eigenvalues $\{\frac{1}{2}\pm e_e\}$. Now, for $\rho_{AE}$ to be dense codeable while $\rho_{AB}$ is not, we want $S(\rho_B)<S(\rho_E)<1$, which reduces to $e_b>e_e$. Using Eqs. (\ref{eq:rhoB}) and (\ref{eq:rhoE}), the condition $e_b>e_e$ becomes
    \begin{eqnarray}
       &&(\lambda_2^2-\lambda_3^2)[-(\lambda_2^2+\lambda_3^2)+\lambda_4^2+1-\lambda_1^2-2\lambda_4^2]>0\nonumber\\
        &&\implies \lambda_0^2(\lambda_2^2-\lambda_3^2)>0 
        \implies |\lambda_2|>|\lambda_3|,
    \end{eqnarray}
where, in the last step, we use $\lambda_0\neq 0$, which refers to the trivial case when the tripartite state is separable across \(A:BE\) cut and hence the proof. 

\end{proof}


\noindent All the three-qubit states satisfying Lemma $1$ may not be suitable for DC swapping. 
As mentioned before, to activate dense codeability in \(AB\) while making \(\mathcal{C}(\rho_{AE})\) classical,  we additionally require to impose a condition that 
$\rho_{AB}$ belongs to the non-ACVENN class of states. In other words,  we now determine the constraints on the parameter space of \(\ket{\psi}_{ABE}\) that must be satisfied for \(\rho_{AB}\notin ACVENN\).\\   

\noindent\textbf{Lemma $2.$(ACVENN conditon for three-qubit pure states):} \textit{For a pure three-qubit state $\ket{\psi_{ABE}}$, its reduced density matrix $\rho_{AB}$ becomes a state in the ACVENN class if and only if the Schmidt parameters of the tripartite state $\ket{\psi_{ABE}}$ follows the following constraint equation: }
\begin{eqnarray}
&&1+4\Bigl(\lambda_2^4+2\lambda_1\lambda_2\lambda_3\lambda_4\cos{\phi}
+\lambda_4^2(-1+\lambda_3^2+\lambda_4^2)\nonumber\\
&&+\lambda_2^2(-1+\lambda_1^2+2\lambda_4^2)=0.\nonumber
\end{eqnarray}

\begin{proof}
    From Lemma 1, $e_b$ and $e_e$ cannot vanish, since in that case, $S(\rho_B) = S(\rho_{AE}) = 1$ and $S(\rho_E) = S(\rho_{AB}) = 1$. The first cannot hold because $\rho_{AE}$ is dense codeable while the second is also not true since $\rho_{AB}$ does not belong to ACVENN. Note that $\rho_{AB}$ belongs to ACVENN only when $S(\rho_E)=1$), i.e., 
\begin{eqnarray}
    &&e_e = 0\nonumber \\   &\implies&1+4\Bigl(\lambda_2^4+2\lambda_1\lambda_2\lambda_3\lambda_4\cos{\phi}+\lambda_4^2(-1+\lambda_3^2+\lambda_4^2)\nonumber\\
    &&+\lambda_2^2(-1+\lambda_1^2+2\lambda_4^2)=0
\label{ACVENN}    
\end{eqnarray}
 holds. This type of states cannot be useful for dense coding swapping.  
\end{proof}

\begin{figure}[h]
    \includegraphics[width=0.9\linewidth]{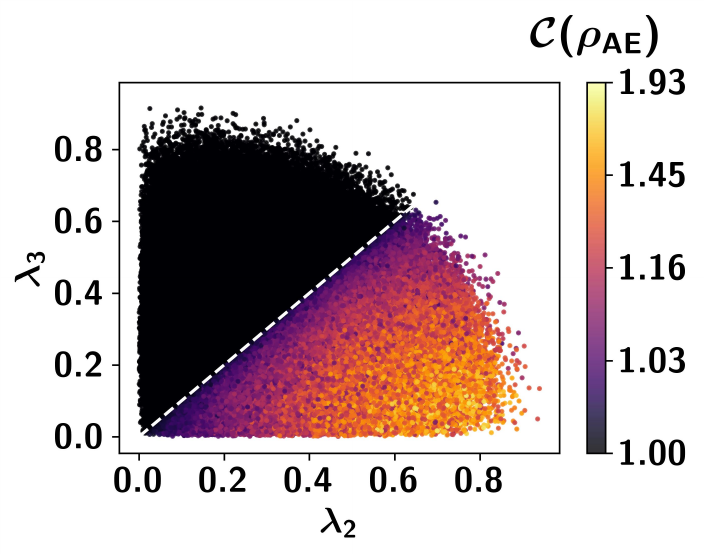}
    \caption{\justifying{{\bf Dense coding capacity of the reduced state $\rho_{AE}$ as a function of the Schmidt coefficients .}
The map plot of \(\mathcal{C}(\rho_{AE})\) for reduced states $\rho_{AE}=\Tr_B[\ket{\psi}\bra{\psi}_{ABE}]$ with respect to $\lambda_2$ (ordinate) and $\lambda_3$  (abscissa) obtained from $5\times10^5$ randomly generated tripartite pure states $\ket{\psi}_{ABE}$.
The black dots represent states that are non-advantageous for dense coding, while the dots with lighter shades correspond to tripartite states with higher dense coding capacity in $\rho_{AE}$. The white dashed line denotes $\lambda_2 = \lambda_3$. The figure shows that the reduced states with $\lambda_2 > \lambda_3$ ensure dense codeability of $\rho_{AE}$. This figure is generated over all set of $\rho_{ABE}$ for which $\rho_{AB} \notin ACVENN$, thus characterizing the set of all resource states for dense coding swapping. All the axes are dimensionless.}}
    \label{fig:PureState}
\end{figure}

\noindent However, the set of three-qubit states becomes a resource if it lies within the set of states that follows both  \textit{Lemma $1$} and the complementary statement of  \textit{Lemma $2$}. This leads us to the following theorem, giving the criteria of state parameters useful for the transformation in Eq. (\ref{eq:switch}).\\ 
 
\noindent\textbf{Theorem 2 :} \textit{A pure three-qubit state can be beneficial  for swapping dense coding ability from \(AE\) to \(AB\) via a global unitary on \(AB\) if and only if the Schmidt coefficients of three-qubits satisfy the conditions $|\lambda_2|>|\lambda_3|$ and \(\lambda_0 \neq 0\) and do not satisfy  Eq. (\ref{ACVENN}).}\\   
 
\noindent For illustration, we Haar randomly  generate  three-qubit pure states for which $|\lambda_2|>|\lambda_3|$, ensuring the dense codeability of $\rho_{AE}$ (see Fig. \ref{fig:PureState}). In addition, we ensure $\rho_{AB} \notin ACVENN$. From this figure, one can visualize Theorem 2. Moreover, we explicitly study some classes of three-qubit states and verify their suitability for DC swapping. \\ 
\noindent \textit{Example.} For the generalized three-qubit \(W\) (gW)  state~\cite{Dur_PRA_2000} $\ket{\psi^{gW}}_{ABE} = \sqrt{a}\ket{001}+\sqrt{b}\ket{010}+\sqrt{1-a-b}\ket{100}$, the corresponding Schmidt form is $\sqrt{1-a-b}\ket{000}+\sqrt{a}\ket{101}+\sqrt{b}\ket{110}$. The non-vanishing Schmidt parameters, in this case,  are found to be $\lambda_2 = \sqrt{a}$, $\lambda_3 = \sqrt{b}$ and $\lambda_0 = \sqrt{1-a-b}$ (see Appendix \ref{A}). For this class of states, the condition given in Lemma 2. reduces to $a=1/2$. Fixing the parameter $a=1/2$, the resulting class of states generates ACVENN states in parties $A$ and \(B\), as the eigenvalues are  $\{0,0,1/2,1/2\}$, independent of the parameter $b$, and have unit Von-Neumann entropy. Therefore, by Theorem $2$,  the conditions that the states are useful for dense coding swapping read as $a\neq \frac{1}{2}$ (non-ACVENN state) and $a>b$, ensuring dense codeability in \(\rho_{AE}\). Eg. choose 
 \(   \ket{\psi_{(1)}^{gW}}_{ABE} = \frac{1}{\sqrt{3}}\ket{001}+ \frac{1}{\sqrt{6}}\ket{010}+ \frac{1}{\sqrt{2}}\ket{100}\), 
whose Schmidt form is $\ket{\psi^{gW}}_{ABE} = \frac{1}{\sqrt{2}}\ket{000}+ \frac{1}{\sqrt{3}}\ket{101}+ \frac{1}{\sqrt{6}}\ket{110}$ with the Schmidt parameters being $\lambda_0 = 1/\sqrt{2}$, $\lambda_2 = 1/\sqrt{3}$, and $\lambda_3 = 1/\sqrt{6}$, while the remaining parameters vanish (see Appendix \ref{A}). According to Theorem $2$, this state serves as a resource for the activation of dense codeability in the subsystem \textit{AB}. This can be verified by evaluating the dense coding capacities of \(\rho_{AB}\)  and \(\rho_{AE}\), which are found to be $1$ bit and $1.26827$ bits, respectively. Moreover, \(\rho_{AB}\)  does not belong to the ACVENN class of states, as the eigenvalues of $\rho_{AB}$ are  $\{\frac{2}{3},\frac{1}{3},0,0\}$ which can be confirmed by computing $ S(\rho_{AB}) = 0.918296 < 1$.





\noindent {\it Dense coding swapping unitary operator.} Let us now illustrate that indeed there exists a two-qubit unitary operator, \(U^{AB}\) which is capable of swapping the quantum advantage of dense coding capacity from \(\rho_{AE}\) to \(\rho_{AB}\).  Specifically, we consider the unitary form \(SU(4)\) of the form~\cite{Vatan_PRA_2004},
\begin{equation}
    U^{AB} = \exp[-i(J_x\sigma_x\otimes\sigma_x+J_y\sigma_y\otimes\sigma_y+J_z\sigma_z\otimes\sigma_z)]
    \label{Eq:EntUnitary}
\end{equation}
where $J_x = 0.785$, $J_y = 0.785$ and $J_z = 0.660$. It can be easily checked that after applying \(U^{AB}\otimes I_E\ket{\psi_{(1)}^{gW}}_{ABE}\), the dense coding capacity of the activated state increases to  $\mathcal{C} (\rho_{AB}) =1.0817$ bits from unity, while  $\rho_{AE}$ becomes non-dense codeable, thereby exhibiting the explicit form of the dense coding swapping unitary operator.

\subsection{Characterizing mixed states as a resource for DC swapping }
\label{subsec:mixed}

Let us now move to the characterization of three-qubit mixed states in terms of resources for DC swapping. 
The approach that we will discuss here is more extensive, involving the Bloch representation of all three-qubit states and finding a condition based on the Bloch parameters that can confirm them as resources. 
Let us express the tripartite mixed state in Fano's form as
\begin{eqnarray}
    &&\rho_{ABE} =\frac{1}{8}(\mathbb{I}_2^{\otimes 3} \nonumber\\
    &&+ \sum_i m_i^{(1)} \sigma_i^{(1)}+ \sum_i m_i^{(2)} \sigma_i^{(2)}+ \sum_i m_i^{(3)} \sigma_i^{(3)}\nonumber\\&+&\sum_{ij} c_{ij}^{(2,3)} \sigma_{ij}^{(2,3)}+ \sum_{ij} c_{ij}^{(1,3)} \sigma_{ij}^{(1,3)}+ \sum_{ij} c_{ij}^{(1,2)} \sigma_{ij}^{(1,2)}\nonumber\\&+& \sum_{ijk} T_{ijk} \sigma_i\otimes \sigma_j\otimes \sigma_k ),
    \label{Eq:tripartite_bloch}
\end{eqnarray}
where $\sigma_i$s ($i\in{1,2,3}$) denote the Pauli matrices. The superscripts on the Pauli matrices specify the subsystems on which the Pauli operators act, while identity operators act on the remaining subsystems. For example, $\sigma_i^{(1)}=\sigma_i\otimes\mathbb{I}\otimes\mathbb{I}$ and $\sigma_{ij}^{(2,3)}=\mathbb{I}\otimes\sigma_i\otimes\sigma_j$, with analogous interpretations for the other terms. In the following discussion, we investigate the minimal structure of the resource states by analyzing the contributions of the local properties $m_i^{(k)} = \Tr(\rho_{ABE}\sigma_i^{(k)})$, \(\forall k=1,2,3\), bipartite correlations $c_{ij}^{(k,l)}= \Tr(\rho_{ABE}\sigma_{i,j}^{(k,l)})$, \(\forall k,l =1,2,3\), and genuine tripartite correlations $T_{ijk} = \Tr (\sigma_i \otimes \sigma_j \otimes \sigma_k \rho_{ABE})$ of the state. Let us first identify the class of states that are not capable of swapping channels for DC. It can be important to determine the class of states essential for DC swapping.\\ 

\noindent {\it Case (a): States having only genuine tripartite correlations.}\\
 
\noindent\textbf{Observation 1. } \textit{ Genuine tripartite correlations do not decide the dense codeability of the reduced bipartite states.}\\
Suppose we consider a tripartite state for which \(m_i^{(k)}\)s and \(c_{ij}^{(k,l)}\)s vanish, i.e., 
\begin{equation}
    \rho_{ABE} = \frac{1}{8}(\mathbb{I}_2\otimes\mathbb{I}_2\otimes\mathbb{I}_2 +\sum_{i,j,k} T_{ijk} \sigma_i\otimes\sigma_j\otimes \sigma_k).
\end{equation}
Both the two-party reduced states $\rho_{AB}$ and $\rho_{AE}$ are the maximally mixed states, implying their non-densecodeablity and they both belong to the ACVENN class of states. Certainly, these states are not resource states for our purpose, the reason being that the only non-trivial term is tripartite genuine correlations, which do not contribute to bipartite correlation and hence, the dense coding capacity of the reduced states.\\

\noindent {\it Case (b): States with only local Bloch vectors.}\\
 
 \noindent\textbf{Observation 2.} We now consider states with only nonvanishing single-site Bloch vectors in \(\rho_{ABE}\),  i.e., tripartite states of the form, \(\rho_{ABE} =\frac{1}{8}\Big(\mathbb{I}_2^{\otimes 3} + \sum_i m_i^{(1)} \sigma_i^{(1)}+ \sum_i m_i^{(2)} \sigma_i^{(2)}+ \sum_i m_i^{(3)} \sigma_i^{(3)}\Big)\)  are irrelevant in characterizing the resource states for our case, since they do not possess entanglement in any bipartitions and hence cannot be dense codeable. 
Although single-site properties like magnetization alone are insufficient to quantify resources required for DC swapping,  they may impose important constraints on the bipartite correlations that determine whether a state can act as a resource or not.\\

\noindent \textit{Case (c): Role of bipartite correlations in DC swapping. }\\

\noindent We therefore adopt an indirect approach: first, we characterize dense codeable and non-dense codeable bipartite states through their correlation matrices, and subsequently apply these characterization criteria to the bipartite reduced states derived from the tripartite systems.\\ 

\noindent Consider an arbitrary two-qubit state, given by
\begin{eqnarray}
&&\rho_{12} ={}\nonumber\\&& \frac{1}{4}(\mathbb{I}_2^{\otimes 2} + \sum_i (m_i \sigma_i\otimes \mathbb{I}_2+ n_i \mathbb{I}_2\otimes \sigma_i) + \sum_{ij} c_{ij} \sigma_i\otimes \sigma_j),\nonumber \\
\label{Eq.bipartite state}
\end{eqnarray}
where $\vec{m}=(m_1,m_2,m_3)^T$ and $\vec{n}=(n_1,n_2,n_3)^T$ denote the local magnetization vectors of parties $1$ and $2$, respectively, and $C = [c_{ij}]_{i,j = \{1,2,3\}}$ represents the bipartite correlation matrix. 
Our goal is to derive a sufficient condition, expressed in terms of the Bloch parameters, for a bipartite state given in Eq. (\ref{Eq.bipartite state}) to be dense codeable. To this end, we employ Pinsker's inequality~\cite{hirota2020pinskerinequality}, which provides a lower bound on the relative entropy between two arbitrary bipartite states,  $D(\rho||\sigma)=\Tr[\rho(\log_2 \rho - \log_2 \sigma)]$  in terms of their trace distance,
\begin{eqnarray}
D(\rho||\sigma) \geq \frac{2}{\ln(2)} T(\rho,\sigma)^2, 
\label{eq:Pinskerineq}
\end{eqnarray}
where
\(T(\rho,\sigma) = \frac{1}{2}||\rho-\sigma||_1\)
is the trace distance between the states $\rho$ and $\sigma$, with $||A||_1 = \Tr[\sqrt{A^\dagger A}]$ being the trace norm of an arbitrary matrix $A$.

\noindent We utilize this inequality to identify a class of dense codeable bipartite states through their correlation coefficients ${c_{ij}}$. Let $\{c_i\}_i$ be the singular values of the correlation matrix $C$. It can then be shown that the state in Eq. (\ref{Eq.bipartite state}) can be rewritten as
\(\tilde{\rho}_{12} = \frac{1}{4}\Big(\mathbb{I}_2^{\otimes 2} + \sum_i m_i \sigma_i\otimes \mathbb{I}_2+ n_i \mathbb{I}_2\otimes \sigma_i )+ \sum_i c_{i} \sigma_i\otimes \sigma_i\Big)\),
which is locally unitarily equivalent to the state given in Eq. (\ref{Eq.bipartite state}). Since the dense coding capacity remains invariant under local unitary transformations, it is sufficient to establish the dense codeability criterion for \(\tilde{\rho}_{12}\). 

It turns out that when \(\vec{m}, \vec{n} \neq 0\) and all \(c_i\)-s are nonvanishing, obtaining the condition in terms of magnetization and correlations is not easy. However, if we restrict the class of states having \(\vec{m} =0\) and others nonvanishing, i.e., the local density matrix $\rho_1 = \Tr_2(\rho_{12})$ is maximally mixed, we arrive at the following theorem. Before presenting that, let us argue that such a class of states can be obtained in a physical set-up. 
Suppose \(A\)  prepares a maximally entangled state across the \(A:A'A"\) bipartition and subsequently transmits the subsystems \(A'\) and \(A"\) to parties \(B\) and \(E\)  through the quantum channels $\Lambda_{A'\rightarrow B}$ and $\Lambda_{A"\rightarrow E}$, respectively. Since the initial state is maximally entangled across the \(A:A'A"\) bipartition, and the channels act locally only on the transmitted subsystems \(A'\) and \(A"\), the reduced density matrix \(\rho_A\) of \(\rho_{ABE}\) remains invariant and maximally mixed, leading to local magnetization vector of $\rho_A$ is vanishing. 


\noindent \textbf{Theorem 3.} \textit{For an arbitrary two-qubit state  with vanishing $\vec{m}$, the sufficient criteria for the states to yield quantum advantage in DC reads as}
\begin{equation}
    \sum_i c_i>2\sqrt{2\ln{2}},
\end{equation}
\textit{where $c_i$s are the singular values of the correlation matrix.}\\

\begin{figure}[t]
    \includegraphics[width=1.0\linewidth]{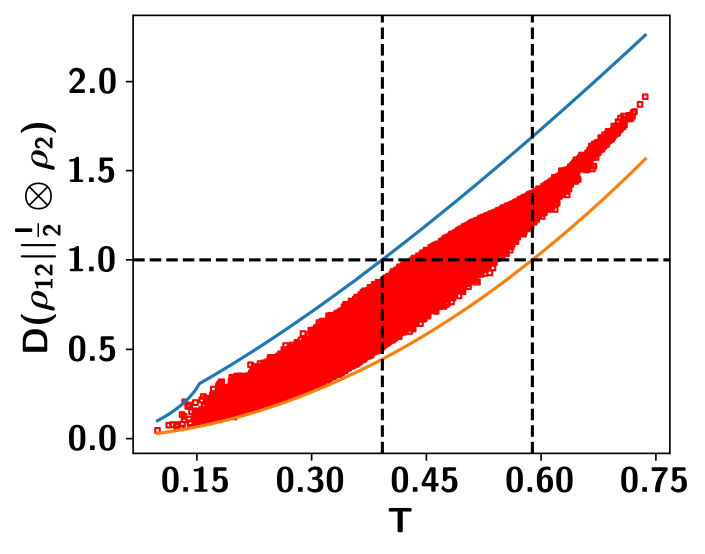}
    \caption{\justifying{ Bounds on relative entropy \(D(\rho_{12}||\frac{\mathbb{I}_2}{2}\otimes \rho_2)\) (ordinate) in terms of trace distance between an arbitrary bipartite state $\rho_{12}$ and $\frac{\mathbb{I}_2}{2}\otimes \rho_2$ (abscissa).
    The figure is generated for $10^5$ rank-$2$, rank-$3$, and rank-$4$ bipartite Haar random mixed states. The blue line is the upper bound given by the expression in Eq.  (\ref{eq:upperbound1}) with $\beta_{opt}$ as $0.152863$. The orange line is the lower bound described by the Pinsker inequality in Eq. (\ref{eq:Pinskerineq}). The dashed vertical line at $T\approx 0.589$ denotes the value of  $T$ above which the states are dense codeable, while the dashed vertical line at $T\approx 0.392$ denotes the value of $T$ below which the states are non-dense codeable. The horizontal line at \(1\) represents the value of the relative entropy above which the state $\rho_{12}$ provides quantum advantage in the dense coding protocol; below it the dense capacities of the states are $1$.  All the axes are dimensionless.}}
    \label{fig:ResourceRegion_Boundaries}
\end{figure}

\begin{proof}
For an arbitrary two-qubit state \(\tilde{\rho}_{12}\) where \(1\) and \(2\) act as the sender and the receiver, respectively, the reduced state at the receiver end is given by
 \(   \rho_2 = \frac{1}{2}(\mathbb{I}_2+\sum_i n_i \sigma_i)\).
In the two-qubit case with \(\vec{m} =0 \), the dense coding capacity reduces to 
\begin{equation}
    \mathcal{C}(\rho_{12}) \coloneqq \max \{1, D(\rho_{12}||\frac{\mathbb{I}_2}{2}\otimes \rho_2)\}.
\end{equation}
 Using the quantum Pinsker inequality in Eq. (\ref{eq:Pinskerineq}), the lower bound on the dense coding capacity can be obtained in terms of the trace norm of the operator $\rho_{12}-\frac{\mathbb{I}_2}{2}\otimes \rho_2$, i.e., $||\rho_{12}-\frac{\mathbb{I}_2}{2}\otimes \rho_2||_1 = \frac{1}{4}||\sum_i c_i \sigma_i\otimes \sigma_i||_1$. The eigenvalues of this operator are $\{\frac{-c_1-c_2-c_3}{4},\frac{c_1+c_2-c_3}{4},\frac{c_1-c_2+c_3}{4},\frac{-c_1+c_2+c_3}{4}\}$. Without loss of generality, one can assume $c_1>c_2>c_3$ and thus the trace distance can be found as 
\begin{eqnarray}
    &&T(\rho_{12},\frac{\mathbb{I}}{2}\otimes \rho_2)\\
    &=& \frac{1}{2}||\rho_{12}-\frac{\mathbb{I}}{2}\otimes \rho_2||_1 =
    \begin{cases}
        \frac{c_1}{2} & \text{if } c_1>c_2+c_3,\\
        \frac{c_1+c_2+c_3}{4} & \text{if } c_1<c_2+c_3.
        \label{indep. depo. Q}\nonumber\\
    \end{cases}
\end{eqnarray}
This allows us to finally reach a sufficient criterion for dense codeability for this class of states as
\begin{eqnarray}
   &&D(\rho_{12}||\frac{\mathbb{I}_2}{2}\otimes \rho_2) \geq \frac{2}{\ln{2}}T(\rho_{12},\frac{\mathbb{I}_2}{2}\otimes \rho_2)^2>1 \nonumber\\
   \implies&& T(\rho_{12},\frac{\mathbb{I}_2}{2}\otimes \rho_2)^2 > \frac{\ln{2}}{2}\nonumber\\
   \implies&& T(\rho_{12},\frac{\mathbb{I}_2}{2}\otimes \rho_2) > \frac{1}{2}\sqrt{2\ln{2}} \\
   \implies&& \begin{cases}
        c_1>\sqrt{2\ln{2}} \approx 1.177 & \text{if } c_1>c_2+c_3\\
        \sum_i c_i>2\sqrt{2\ln{2}} \approx 2.355& \text{if } c_1<c_2+c_3 \nonumber
        \label{eq.DCcond}\\
    \end{cases}
\end{eqnarray}
The proof holds only for those cases when $c_1<c_2+c_3$ although $c_i\leq 1$ implies that we cannot find a sufficient condition when $c_1>c_2+c_3$. The lower bound on the relative entropy $D(\rho_{12}||\frac{\mathbb{I}_2}{2}\otimes \rho_2)$ equals $1$, or equivalently the conditional von-Neumann entropy $S(1|2)$ vanishes at $T(\rho_{12},\frac{\mathbb{I}_2}{2}\otimes \rho_2) = \frac{1}{2}\sqrt{2\ln{2}} \approx 0.589$ (see Fig. \ref{fig:ResourceRegion_Boundaries}).
\end{proof}

\noindent Let us now begin to derive a sufficient condition for non-dense codeability in terms of the parameters of a two-qubit density matrix with \(\vec{m} =0\). To approach this, we use an upper bound on \(D(\rho||\sigma)\), i.e., \(D(\rho||\sigma) \leq f(\beta, T)\) \cite{Audenart_JMP_2005}, with  

\begin{eqnarray}
    &&f(\beta, T) = \nonumber\\&&
        (\beta+T) \log_2(\frac{\beta+T}{\beta}),~~ \text{if } \beta\leq T\leq 1-\beta{}\nonumber\\
       && (\beta+T) \log_2(\frac{\beta+T}{\beta}) + (\beta-T) \log_2(\frac{\beta-T}{\beta}),  \text{if } T< \beta.\nonumber \\
    \label{eq:upperbound1}
\end{eqnarray}
Here, $\beta$ is the minimum eigenvalue of $\sigma$ ($\lambda_{min}(\sigma)$), and $T$ denotes the trace distance between the states $\rho$ and $\sigma$. Like the previous proof, we consider $\rho=\rho_{12}$ while $\sigma= \mathbb{I}/2\otimes \rho_2$ such that $D(\rho||\sigma)$ becomes the dense coding capacity.
The function $f(\beta,T)$ decreases monotonically with $\beta$ for any fixed value of $T$. Furthermore, our numerical analysis over $10^5$ generated Haar random states indicates that, for a fixed value of $T = T_0$, a value of $\beta$ exists ($\beta^*(T_0)$) such that the function attains its minimum while still being greater than $\mathcal{C}(\rho_{12})$ of all the states $\rho_{12}$ generated with the same value of $T = T_0$. Therefore, to obtain a numerical upper bound, we determine, for a fixed $T=T_0$, the largest value of $\beta$ corresponding to the minimum of $f(\beta,T_0)$ while satisfying $f(\beta^*(T_0), T_0) \geq \mathcal{C}^*_{12}(T_0)$, where $\mathcal{C}^*_{12}(T_0) = \max\{\mathcal{C}(\rho_{12})|T(\rho_{12},\mathbb{I}/2\otimes \rho_2)=T_0\}$. There may exist some other value $T=T_1\neq T_0$ such that $f(\beta^*(T_0), T_0)<\mathcal{C}^*_{12}(T_1)\leq f(\beta^*(T_1), T_1)$. Subsequently, we find the minimum of the optimal $\beta$-s over all possible values of $T$. The resulting value, denoted by $\beta_{opt}$, ensures that $f(\beta_{opt},T)>\mathcal{C}_{12}$ for all numerically generated values of $\beta$ and $T$, and hence for all numerically generated states. This optimization can be expressed as follows:
\begin{eqnarray}
    &&\beta^*(T) = \arg\min_\beta f(\beta,T)\text{  for fixed $T$}\nonumber\\
    \text{subject to}&&\nonumber\\
    && f(\beta,T) \geq \mathcal{C}^*_{12}(T) 
    \label{eq:opti}
\end{eqnarray}
\begin{eqnarray}
    \beta_{opt} = \min_{T} \beta^*(T)
    \label{eq:constraint}
\end{eqnarray}
This method generates $\beta_{opt}$ to be $0.152863$, where we execute the optimization over $10^5$ Haar-random states. \\

\noindent We are now ready to present 
the sufficient condition for a two-qubit arbitrary density matrix with \(\vec{m} =0\) to be non-dense codable.\\ 

\noindent \textbf{Proposition 1.}\textit{ Given an arbitrary two-qubit state with vanishing $\vec{m}$, the non-dense codeablity of the state is guaranteed when}
\begin{eqnarray}
   \sum_i c_i \leq 1.569.
   \label{eq:cupper}
\end{eqnarray}
\begin{proof}
    
As established earlier, the trace distance between the state $\tilde{\rho}_{12}$ and  $\frac{\mathbb{I}_2}{2}\otimes \rho_2$ is equal to one-fourth of the sum of the singular values of their correlation matrix, i.e., $\frac{\sum_i c_i}{4}$, provided \(c_1< c_2 + c_3\). Numerical optimization yields the value $\beta_{opt}=0.152863$. From Fig.~\ref{fig:ResourceRegion_Boundaries}, it is observed that the function $f(T,\beta_{opt})$ drops below unity for  $T >\beta_{opt} = 0.152863$. Consequently, we investigate the parameter regime satisfying $(\beta_{opt}+T)\log_2{\frac{\beta_{opt}+T}{\beta_{opt}}} \leq 1$. Although this equation does not admit a closed-form solution, numerical evaluation shows that it reduces to the condition $T\leq T_{opt} = 0.392278$. Substituting the explicit expression of $T$, we arrive at Eq. (\ref{eq:cupper}), i.e., the sum of singular values of the bipartite correlation for this class of states is upper bounded by \(1.569\).
\end{proof}


\noindent \textbf{Note:} We emphasize that the parameter $\beta$ depends on the choice of $\vec{n}$. Therefore, in order to numerically formulate a one-parameter criterion for non-dense codeability, it is necessary to identify the vector $\vec{n}_{opt}$ corresponding to $\beta_{opt}$ together with the optimal threshold $T_{opt}$. For the ensemble of $10^6$ Haar-random states generated in our analysis, all the numerically generated states have $\beta>\beta_{opt}$; thus,the resulting condition is found to be independent of $\beta$. In this sense, the bound should be regarded as numerical. More generally, one may derive alternative sufficient conditions for any $\beta'>\beta_{opt}$. Such conditions, however, necessarily become biparametric, requiring simultaneous constraints of the form $\beta>\beta'$ and $T<T'$. The functional relation between $\beta'$ and $T'$ is obtained by solving the equation $(\beta'+T')\log_2{\frac{\beta'+T'}{\beta'}} = 1$. This now enables us to picturize the geometric picture of our resource states.\\

\noindent\textbf{Corollary:} \textit{The sufficient condition for the three-qubit arbitrary state in Eq. (~\ref{Eq:tripartite_bloch}), with the entropy of its reduced state $\rho_{AB}$ less than unity, to qualify as a resource for DC swapping can be expressed as }
\begin{enumerate}
    \item $\sum_i c_i^{(1,3)}>2.335$
    \item $\sum_i c_i^{(1,2)} \leq 1.569$
\end{enumerate}
\textit{where $c_i^{(1,3)}$-s and $c_i^{(1,2)}$-s are the singular values of the bipartite correlation matrices $c^{(1,3)}$ and $c^{(1,2)}$  of \(\rho_{AB}\) and \(\rho_{AE}\) respectively.}\\

\noindent \textbf{Remark.} This characterization can be easily extended to multipartite scenario, by replacing $c^{(1,3)}$ by the bipartite correlation matrix of the sender-malicious receiver pair, while replacing $c^{(1,2)}$ by the bipartite correlation matrix of the sender-target receiver pair.

\section{DC swapping resources beyond three-qubits }
\label{sec:morequbitandGGM}

Upto now, we have found the useful three-qubit pure states for DC swapping in terms of Schmidt coefficients. We now extend our investigation to systems comprising three or more qubits, to characterize their suitability for DC swapping in terms of their quantum correlation properties, particularly multipartite entanglement, which serves as a key resource in many quantum information processing tasks. 

Analyzing multipartite Haar-random states that are useful for the dense coding swapping task is troublesome; as the number of parties increases, the number of appropriate states reduces drastically.
\begin{figure}[h]
    \includegraphics[width=1.0\linewidth]{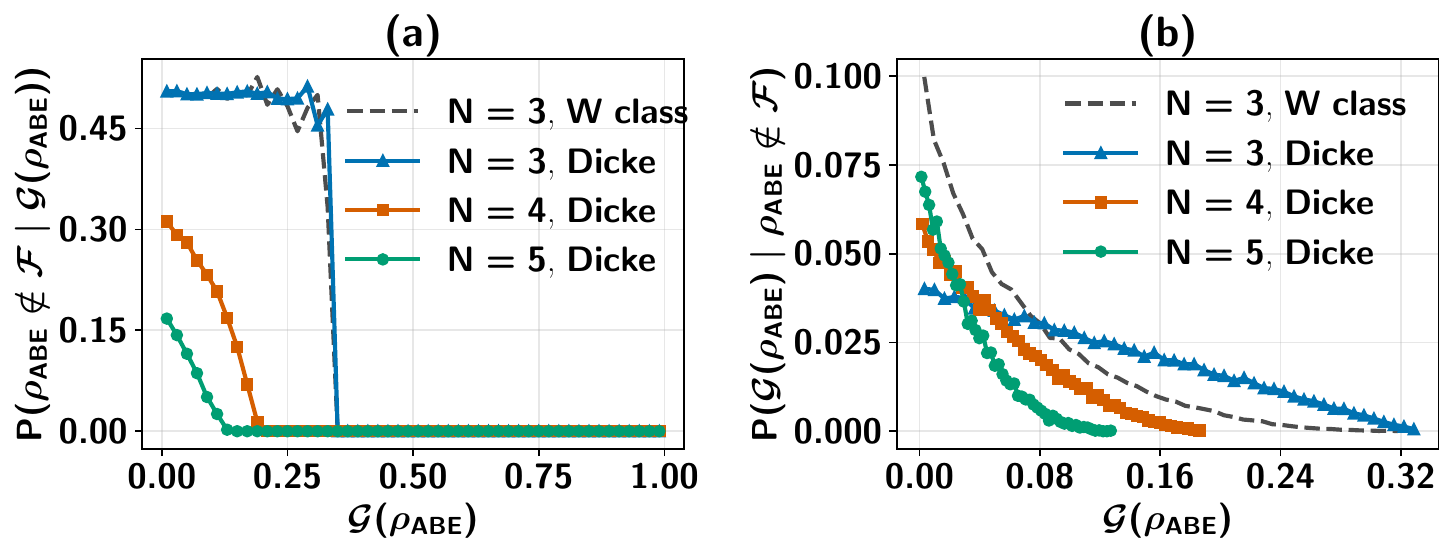}
    \caption{\justifying{{\bf Role of genuine multiparty entanglement in dense coding swapping task.} (a) The probability distribution of a state being a resource state as a function of the generalized geometric measure (GGM).  (b) The probability distribution of the GGM of the initial shared state, conditioned on the state being a resource. The triangles, squares, and circles correspond to three-, four-, and five-qubit Dicke states with a single excitation, respectively. The (grey) dashed line represents the three-qubit W-class states. All axes are dimensionless.
    }}
    \label{fig:ggm-res}
\end{figure}
For example, by generating $10^7$ Haar-random states, we find that the percentage of states that are suitable is $100\%$ in the case of tripartite states, although it reduces to only $3.15\%$ for the four-qubit states and almost none for the five-qubit states. This is consistent with the fact that for a $N$-party pure state of dimension $d_Ad_B$, the average entropy of a subsystem of dimension $d_A$ can be shown to be \cite{Page_PRL_1993}
\begin{eqnarray}
    \langle S_{d_A,d_B}\rangle = \sum_{k=d_B+1}^{d_Ad_B} \frac{1}{k} - \frac{d_A-1}{2d_B},
\end{eqnarray}
where \(d_B\) is the dimension of the \(B\) subsystem. The average entropy given above tends to $\log_2 d_A$ as one increases the dimension of the subsystem \(B\). Thus, asymptotically, i.e., for $N\rightarrow \infty$, the desired reduced states tend to exhibit more mixedness and have higher entropy, resulting in ACVENN states, which cannot be useful for the dense coding swapping task \cite{Kendon_PRA_2002}. This is the reason we choose a subspace of the general Haar random states, especially the family of Dicke states~\cite{Dicke,Bergmann_2013,Asutosh_PLA_2017,Bernd_PRL_2014,Chiuri_PRL_2012}, so that non-ACVENN states occur with significant probability. Although the Dicke states have a set of measure zero, their correlation lives in the bipartitions, which qualifies them as suitable for our analysis. 
The Dicke states with \(r\) excitation can be written as 
\begin{eqnarray}
    \ket{\psi_N^r} = \sum_k a_k \mathcal{P}_k[\ket{0}^{\otimes N-r} \ket{1}^{\otimes r}]
\end{eqnarray}
with $a_k = \alpha_k + i \beta_k$ being the complex numbers ($\alpha_k$ and $\beta_k$ are real numbers with \(i = \sqrt{-1}\)) such that $\sum_k|a_k|^2 = 1$ and the summation is over all permutations $\{\mathcal{P}_k\}$ of the product states $\ket{0}^{\otimes N-r} \ket{1}^{\otimes r}$ with $r$ qubits in the excited state $\ket{1}$ and the rest are in the ground state $\ket{0}$.
We consider Dicke states with three-, four-, and five-qubits with one excitation for our analysis. This study is bidirectional in the sense that it (1) depicts the probability distribution of resource states across the range of the shared state's initial entanglement, i.e., the distribution probability that the state is a resource given the state's initial genuine multiparty entanglement ($P(\rho_{ABE}\notin\mathcal{F}|\mathcal{G}(\rho_{ABE}))$) and also illustrates (2) the distribution of initial multiparty entanglement given that the state is a resource, i.e., the probability of the state's initial genuine multiparty entanglement given that the state is a resource ($P(\mathcal{G}(\rho_{ABE})|\rho_{ABE}\notin\mathcal{F})$), where \(\mathcal{G}\) is the genuine multipartite entanglement measure. The numerical analysis is performed by generating $10^6$ Haar-random Dicke states directly inside the Dicke subspace by randomly drawing coefficients from normal distributions. For example, for four-qubit single-excitation Dicke states, we generate the coefficients $\alpha_k+i\beta_k$ corresponding to the basis state with excitation at the $i^{th}$ position ($i\in \{1,2,3,4\}$) by randomly drawing $\alpha_k$ and $\beta_k$ from a normal distribution, i.e., a Gaussian distribution with mean zero and unit variance \cite{Ratul_PRA_2020}. We also compute the generalized geometric measure (GGM), which quantifies the genuine multiparty entanglement of an $N$-party pure quantum state~ \cite{Wei_PRA_2003,Aditiggm_PRA_2010, Tamoghna_PRA_2016}. It is defined as an optimized distance of the given state from the set of all states that are not genuinely multiparty entangled. Mathematically,
\begin{eqnarray}
    \mathcal{G}(\ket{\psi_N}) = 1 - \max_{\ket{\chi}} \Lambda_{max}^2({\ket{\psi_N}})
\end{eqnarray}
where $\Lambda_{\max}(\ket{\psi_N})= |\langle \chi | \psi_N\rangle|$ and the maximization is performed over the set containing all $\ket{\chi}$s that are not genuine multiparty entangled. An equivalent computable form of GGM is~\cite{Aditiggm_PRA_2010}
\begin{eqnarray}
    &&\mathcal{G}(\ket{\psi_N})\\
    &=& 1 - \max\{\lambda^2_{I:L}| I \cup L = \{A_1,A_2,\cdots,A_N\},I\cap L = \emptyset \}, \nonumber
\end{eqnarray}
where $A_i$s denote the subsystems of the $N$-party state $\ket{\psi_N}$ and the maximization is performed over all such bipartitions.

Our analysis reveals that genuine multiparty entanglement does not help the DC swapping task. In fact, the resource states generally have very low GGM values. This is due to the fact that the DC swapping deals with dense codeability between two parties.  We also observe a trade-off relation between GGM and the number of qubits: as the number of qubits increases, resource states tend to concentrate on the lower GGM values.
Further, we notice that the probability of obtaining a state that is a resource is higher when the state has a lower GGM value. For a lesser number of qubits, resource states can be found at relatively higher GGM values. It is also interesting to notice that resource states are less and less frequent as one increases the number of parties (see Fig. \ref{fig:ggm-res}).

{\it Symmetric multipartite states -- not useful for DC swapping.} Clearly, the states that are symmetric under permutation of parties cannot act as a resource for the transformation in Eq. (\ref{eq:switch}), i.e., they are not advantageous for a DC swapping protocol due to the exclusion principle of DC~\cite{prabhu2012exclusion}. Prominent examples include both the generalized GHZ (gGHZ) state~\cite{Greenberger_arXiv_2007,Lomonaco_arXiv_2004}, 
\begin{figure}[h]
    \includegraphics[width=1.0\linewidth]{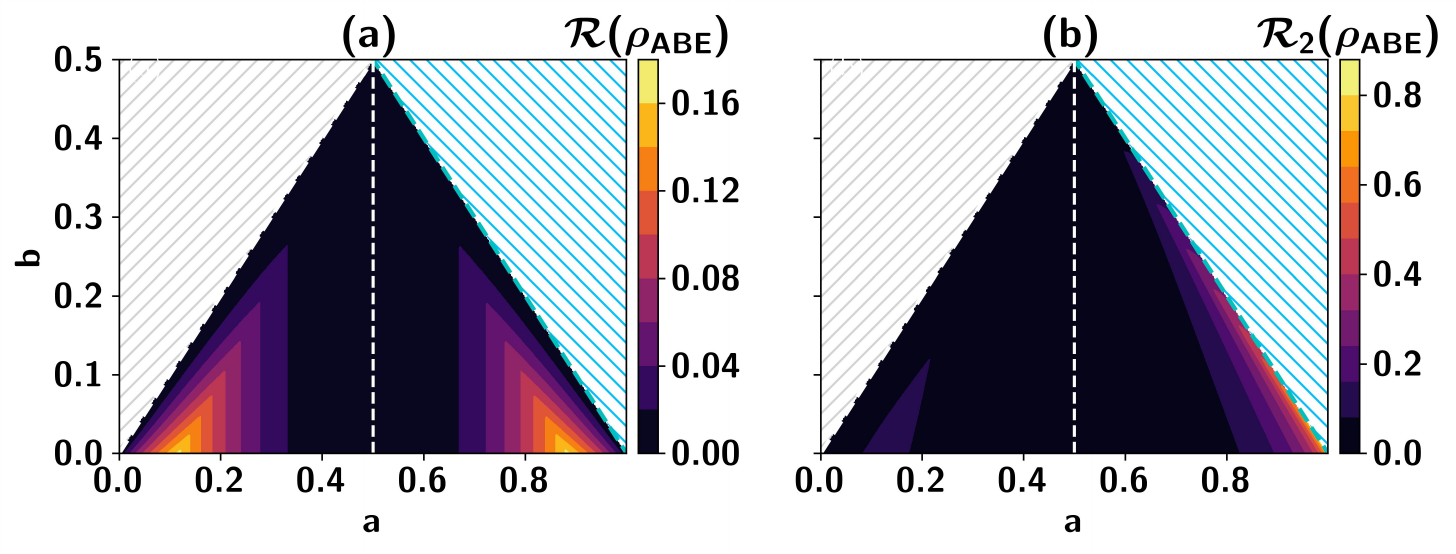}
    \caption{\justifying{{\bf Tolerance of the generalized W states across the parameter space $a$ and $b$ against noise.}
   (a) Map plot for the tolerance of the resource states, \(\mathcal{R}(\rho_{ABE})\) against \(a\) and \(b\) of the gW states. (b) The same when the gW state is admixed with orthogonal colored noise described in the text. The grey-hatched region on the left corresponds to non-resource states satisfying $a<b$, whereas the blue-hatched region on the right represents states with non-real coefficients, for which $1-a-b<0$. The white dashed line at $a=0.5$ marks the set of states whose reduced state $\rho_{AB}$ belongs to the ACVENN class. Within the physically admissible resource region (the triangular domain), brighter shades indicate states exhibiting greater tolerance against white noise. Values of tolerance that are less than $10^{-3}$ are regarded as negligible and are therefore treated as \(\mathcal{R}(\rho_{ABE})=0\). All axes are dimensionless.}}
    \label{fig:Robustness}
\end{figure}
(\(\ket{gGHZ} = (\alpha \ket{00\ldots 0} + \beta \ket{11\ldots1})\)) with \(\alpha\) and \(\beta\) being complex numbers with proper normalization and the \(N\)-qubit \(W\) state~\cite{Dur_PRA_2000, Aditi_PRA_2003}, \(\ket{W}  \equiv |\psi_N^{(1)}\rangle = \frac{1}{\sqrt{N}}\sum \mathcal{P}_i\ket{10\ldots0}\) where \(\mathcal{P}_i\) permutes the product state with  \(1\) excitation. However, asymmetric states are also not guaranteed to be beneficial for the activation scheme, as illustrated in Sec. \ref{sec:charac} and in this section. 
\\

\section{Tolerance of pure states against white and colored noise}
\label{sec:robustness}

Noise can have a debilitating effect on the ability of a quantum state to act as a resource
which turns out to be one of the main challenges in implementing quantum information processing tasks. In the present section, we study the tolerance of useful class of tripartite pure states identified in Sec. \ref{sec:charac} against white noise. 
 We quantify the tolerance of the state to noise as the minimum noise admixture probability \(s\) at which the state looses its resourceful character for DC swapping. Mathematically,
\begin{eqnarray}
     \mathcal{R}(\rho) = \min \{\exists\, s\geq 0| \frac{\rho_{ABE} + s \mathbb{I}/d}{1+s}\in \mathcal{F}\}.
\end{eqnarray}
In other words, it tells that the state remains suitable for dense coding swapping even after admixing the maximum amount of white noise to the three-qubit pure state \(\ket{\psi}_{ABE}\).\\  
In this analysis, we choose those class of initial pure resource states, such as the generalized W state $\ket{\psi^{gW}}_{ABE}$ (with \(a, b\geq 0)\) and \(a + b \leq 1\)), which are shown to be useful for DC swapping in the absence of noise. For investigating its tolerance against noise, we consider the dense coding swapping capability of \(\rho^{gW}_{ABE}= \frac{\ketbra{\psi^{gW}}_{ABE} + s\mathbb{I}/8}{1+s}\). As shown in Sec. \ref{sec:charac}, the dense codeability of \(\rho_{AE}\) of \(\ketbra{\psi^{gW}}_{ABE}\) can be transferred to \(\rho_{AB}\) when \(a\neq 1/2\) and \(a>b\) of the gW state. We observe that for small values of \(b\), there exists a wide range of \(a\) for which the state \(\rho^{gW}_{ABE}\) is good for DC swapping. E.g., we find that 
\(\mathcal{R}(\rho^{gW}_{ABE}) \approx 0.165\), when \(a\) either is close to \(\approx 0.11\) or \(\approx 0.88\) and \(b\) is close to zero (see Fig. \ref{fig:Robustness}(a) for details). As expected from the analysis of the pure state, when \(a\) is close to \(1/2\), it is less robust as compared to the region with \(a\) close to \(0\) and \(1\). There exists  a threshold value \(b\) above which the state is not resilient to noise. Note, however, that
the entire numerical analysis is carried out with the precision \(\mathcal{O}(10^{-3})\), i.e., the tolerance is set to \(0\)  when \(s < 10^{-3}\). 

\noindent We also extend the analysis of tolerance for orthogonal colored noise, which can be defined as
\begin{eqnarray}
     \mathcal{R}_2(\rho_{ABE}) = \min \{\exists\, s\geq 0| \frac{\rho_{ABE} + s \rho^{\perp}_{ABE}}{1+s}\in \mathcal{F}\},
\end{eqnarray}
where $\rho^{\perp}$ is defined by the orthogonality condition $\Tr[\rho \rho^{\perp}]=0$. 
For the gW state, the orthogonal colored noise is taken as $\ket{\psi^{gW}}^{\perp} = \sqrt{\frac{b}{a+b}}\ket{001}-\sqrt{\frac{a}{a+b}}\ket{010}$.  For this type of noise,  a different tolerance pattern emerges for the generalized W states. Firstly, under this kind of colored noise, \(\mathcal{R}_2(\rho_{ABE})\) is no more symmetric with respect to \(a=1/2\). Secondly, greater tolerance to noise is achieved when $a$ is high, close to unity, and \(b\) is small (see Fig. \ref{fig:Robustness}(b)) in comparison to gW states with low \(a\) values.  Interestingly, however, \(\mathcal{R}_2(\rho_{ABE})\) can reach much higher values than \(\mathcal{R}(\rho_{ABE})\) obtained via white noise, thereby demonstrating more noise resistance under colored noise than under white noise.


\section{Characterizing unitaries capable of dense coding swapping}
\label{sec:unitary}

 We now address the question -- "Can the parameters in two-qubit unitaries in Eq. (\ref{Eq:EntUnitary})  follow certain rules, suitable for DC swapping?"  We can answer this query affirmatively for the useful classes of states discussed in Secs. \ref{sec:charac}  although it is mathematically tedious to address for arbitrary states.  Note that the parameter-characterization of unitaries in Eq. (\ref{Eq:EntUnitary})   is enough since the dense coding capacity remains invariant under local unitaries involved in the decomposition of  \(SU(4)\).


\noindent Let us begin by characterizing the unitaries that allow Alice and Bob to activate maximum dense codeability in $\rho_{AB}$ by making  \(\rho_{AE}\) NDC when the initial shared state is the generalized W state.\\

\noindent \textbf{Theorem 3.} \textit{Considering the  generalized W state, $\ket{\psi^{gW}}_{ABE} = \sqrt{a}\ket{001}+\sqrt{b}\ket{010}+\sqrt{1-a-b}\ket{100}$ with real coefficients, i.e., $a\geq 0$, $b\geq 0$ and $a+b\leq 1$ (corresponding Schmidt form $\ket{\psi^{gW}}_{ABE} = \sqrt{1-a-b}\ket{000}+\sqrt{a}\ket{101}+\sqrt{b}\ket{110}$),
 we find the conditions on optimal unitary capable of swapping dense codeability from \(\rho_{AE}\) to \(\rho_{AB}\)}: 
 (1) $\mathcal{C}\Big(\Tr_E \Big[U^{AB}\ketbra{\psi^{gW}}{\psi^{gW}}_{ABE}(U^{AB})^{\dagger}\Big]\Big)$ \textit{is independent of $J_z$}; and 
(2) \textit{The optimal unitary $U^{AB}_*$, for which $\max_{U^{AB}} \mathcal{C}\Big(\Tr_E \Big[U^{AB}\ketbra{\psi^{gW}}{\psi^{gW}}_{ABE}(U^{AB})^{\dagger}\Big]\Big)$ is achieved, is characterized by}
    \begin{eqnarray}
        \tan{2 J^*_x} \tan{2 J^*_y} = \frac{1-2b}{1-2a-2b},
        \label{Eq:OptimumU_for_gW}
    \end{eqnarray}
    \textit{where $J^*_x$ and $J^*_y$ are the parameters of the optimal unitary $U^*$}.
\begin{proof}
    We outline the proof in the following steps:\\
Step 1.  We commence by applying $U^{AB}\otimes\mathbb{I}_2^{E}$ on $\ket{\psi^{gW}}_{ABE}$:
    \begin{eqnarray}
        &&\ket{\phi}_{ABE}\equiv U^{AB}\otimes\mathbb{I}_2^{E}(\ket{\psi^{gW}}_{ABE})\nonumber\\
        &=& e^{-iJ_z}\sqrt{a} \cos(J_x-J_y) \ket{001}\nonumber\\ &+& e^{iJ_z}(\sqrt{b} \cos(J_x+J_y) - i\sqrt{1-a-b}\sin(J_x+J_y))\ket{010} \nonumber\\&+& e^{iJ_z}(\sqrt{1-a-b} \cos(J_x+J_y) - i\sqrt{b}\sin(J_x+J_y))\ket{100} \nonumber\\&-& i e^{-iJ_z}\sqrt{a} \sin(J_x-J_y) \ket{111}.
    \end{eqnarray}
\begin{widetext}
By tracing out the subsystem $E$, the resulting state  $\sigma_{AB}^{gW}$ is given by 
\begin{eqnarray}
        &&\sigma^{gW}_{AB} = \Tr_E\Big[{\ket{\phi}_{ABE}\bra{\phi}_{ABE}}\Big]\nonumber\\
        &=& U^{AB} \rho_{AB}^{gW} (U^{AB})^{\dagger}\nonumber\\
        &=&
        \begin{bmatrix}
            a \cos^2(J_{x-y}) & 0 & 0 & \frac{i}{2} a \sin(2J_{x-y})\\
            0 & b \cos^2(J_{x+y})+(1-a-b)\sin^2(J_{x+y}) & (A-iB)(C+ iD) & 0 \\
            0 & (A+iB)(C-iD) & b \sin^2(J_{x+y})+(1-a-b)\cos^2(J_{x+y}) & 0\\
            -\frac{i}{2} a \sin(2(J_{x-y})) & 0 & 0 &  a \sin^2(J_{x-y}),
        \end{bmatrix}\nonumber\\
        \label{Eq:UonAB}
    \end{eqnarray}
    where $A \pm iB=\sqrt{b}\cos(J_{x+y})\pm i\sqrt{1-a-b}\sin(J_{x+y})$, $C \pm iD=\sqrt{1-a-b}\cos(J_{x+y})\pm i\sqrt{b}\sin(J_{x+y})$ $J_{x\pm y}=J_x \pm J_y$.
    \end{widetext}
   Clearly, it is independent of the parameter $J_z$, implying its DC capacity does not depend on $J_z$.

\noindent  Step 2.  Since the reduced density matrix after applying the unitary  $\sigma^{gW}_{AB}$ does not involve the parameter $J_z$, we can safely take $J_z \in [0,\pi/2]$ for the optimal unitary, while the optimality depends on the parameters $J_x$ and $J_y$. Moreover, we notice that the state $\sigma^{gW}_{AB}$ in Eq.(\ref{Eq:UonAB}) is an X-state and its eigenvalues are $\{0,0,a,1-a\}$. 
\begin{figure}[h]
    \includegraphics[width=1.0\linewidth]{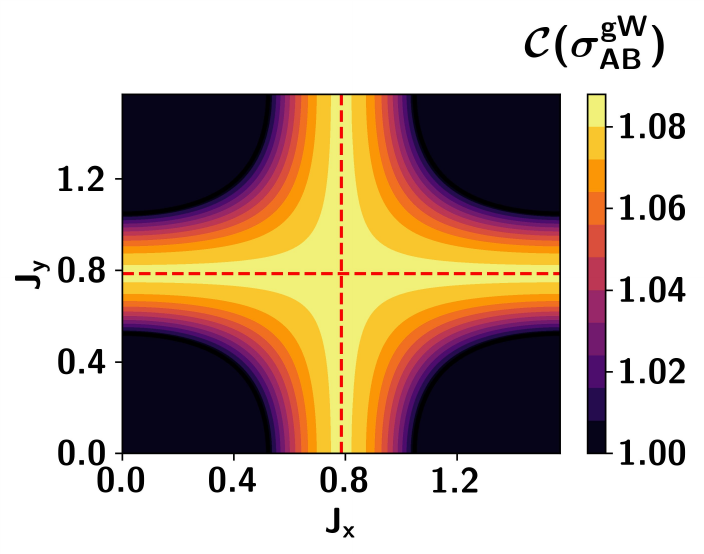}
    \caption{\justifying{{\bf Parameter space specifying the region of dense coding swapping unitaries.} The plot illustrates the variation of the dense coding capacity of the three-qubit gW state defined in \ref{sec:charac} example, with $a=\frac{1}{3}$ and $b=\frac{1}{6}$, after the application of a DC swapping unitary parameterized by $J_x$, $J_y$, and $J_z$ in Eq. (\ref{Eq:EntUnitary}). The parameter $J_z$ is redundant in determining  $\mathcal{C}(\sigma^{gW}_{AB})$. The parameters $J_x$ and $J_y$ are varied over the range $[0,\pi/2]$ with a step size of $0.001$. The red dashed lines at $J_x = \pi/4$ and $J_y = \pi/4$ denote the family of optimal dense coding swapping unitaries for the considered state. The black-shaded region represents parameter combinations that cannot function as DC swapping unitaries, whereas the lighter-shaded region corresponds to unitary parameters that swaps the dense codeability from \(AE\) to \(AB\), resulting in a higher dense coding capacity in the \(AB\) subsystem. All axes are dimensionless.
}
}
    \label{fig:max_unitaryRegion}
\end{figure}

The reduced state $\sigma^{gW}_B = \Tr_A[\sigma^{gW}_{AB}]$ is diagonal in the computational basis and its eigenvalues are $\{a\cos^2(J_x-J_y)+b\sin^2(J_x+J_y)+(1-a-b)\cos^2(J_x+J_y), a\sin^2(J_x-J_y)+b\cos^2(J_x+J_y)+(1-a-b)\sin^2(J_x+J_y)\}$. For a given gW state, a and b are fixed, which implies that the entropy $S(\sigma^{gW}_{AB})$ is fixed. The maximum dense coding capacity achievable is thus obtained by maximizing $S(\sigma^{gW}_{B})$. In general, this happens when the absolute value of the difference between eigenvalues of $\sigma^{gW}_{B}$ is minimum. One can see that if we set $J_x = 0$ and $J_y = \frac{\pi}{4}$, the difference between the eigenvalues vanishes, i.e., the eigenvalues are the same and equal to $\frac{1}{2}$, maximizing the entropy of the state $\sigma^{gW}_{B}$. So, the optimal unitary is obtained when the eigenvalues are equal. In this case, the capacity depends only on $a$ and therefore,  $\mathcal{C}(\sigma^{gW}_{AB}) = 2-H(\{a,1-a\})$, with $H(\cdot)$ being the Shannon entropy defined as $H(\{p_i\}) = -\sum_i p_i \log_2 p_i$. For $a=\frac{1}{3}$ and $b = \frac{1}{6}$, the maximum capacity of $\sigma^{gW}_{AB}$ that can be achieved after applying the optimal unitary is $2-H(\{1/3,2/3\}) = 1.0817$ bits, as shown in Sec. \ref{sec:charac} example.

\begin{figure}[t]
    \includegraphics[width=1.0\linewidth]{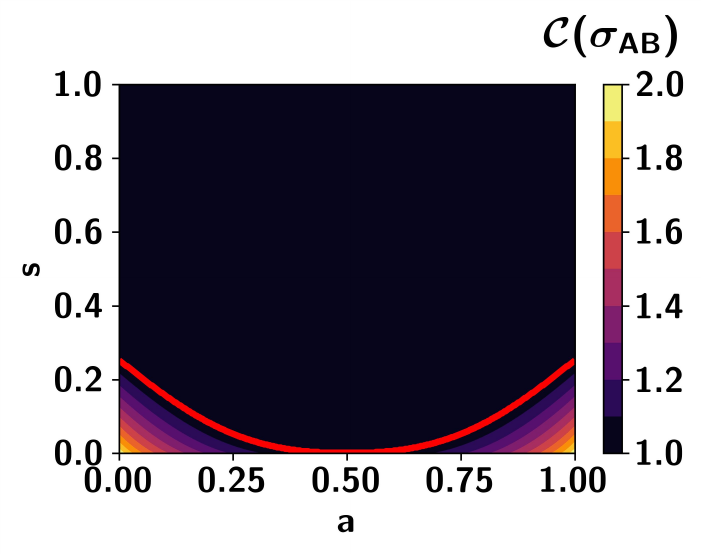}
    \caption{\justifying{ Classification of the parameter space of the white-noise admixed generalized W state according to its capability to generate dense codeability in the reduced state $\sigma_{AB}$ following the application of the optimal unitary. The dense coding capacity of $\sigma_{AB}$ given by Eq. (\ref{Eq:Cap_whitenoise}) is plotted across the initial state's parameter space $(a,s)$. Regions with brighter shades denotes the region in the parameter space $(a,s)$ for which the optimal unitary can swap the dense codeability to \(AB\) from \(AE\). The bright red line (solid) denotes the boundary line where the quantum advantage cease to exist. All axes are dimensionless.
    }}
    \label{fig:DC_activationRegion_noisy}
\end{figure}
It is easy to see that optimal unitaries generating dense codeability in the state $\sigma^{gW}_{AB}$ except for the state when $a=0.5$ which belongs to the ACVENN class of states.
Setting the eigenvalues to be equal, we obtain the condition on Eq.(~\ref{Eq:OptimumU_for_gW})   (see  Appendix \ref{A1}).
\end{proof}

\noindent To illustrate the conditions on parameters in the two-qubit unitaries in Eq. (\ref{Eq:OptimumU_for_gW}), we plot the dense coding capacity of the particular state \(\sigma_{AB}^{gW}\) given in Sec.\ref{sec:charac} example with $a=\frac{1}{3}$ and $b=\frac{1}{6}$ in Fig. \ref{fig:max_unitaryRegion}. The figure shows that when \(J_x =J_y =\pi/4 \approx 0.785\) and when one of them vanishes and the other is \(\pi/4\) in \(U^{AB}\), \(\mathcal{C}(\sigma_{AB}^{gW})\) achieves the maximum value, which can be proved from Eq.(\ref{Eq:OptimumU_for_gW}) (see red dashed line in Fig. \ref{fig:max_unitaryRegion}). \\

\noindent Let us now show that such a characterization of unitaries remains true even when the gW state is admixed with white noise, i.e., for this class of noisy gW states, again conditions 1 and 2 derived in Theorem 3 hold.  Specifically, we obtain the following theorem:\\

\noindent \textbf{Theorem 4.} \textit{For the white noise admixed generalized W state $\rho_{ABE} = (1-s)\ketbra{\psi^{gW}}{\psi^{gW}}+s\frac{\mathbb{I}_8}{8}$, the parameters in the unitary follow the same conditions as in Theorem 3.} 

The proof is given in Appendix \ref{A2}. The DC capacity after applying the unitaries in the parameter space is depicted in Fig. \ref{fig:DC_activationRegion_noisy}. The maximal unitary operators are not always able to generate dense codeability in $\sigma_{AB}^{gW}$  across the entire parameter space of \(a\),  and \(s\) (it is independent of \(b\)). Specifically, they fail to do so for states belonging to the ACVENN class as shown in Fig. \ref{fig:DC_activationRegion_noisy} for high values of \(s\). 



\noindent  This analysis can further be extended when the state \( \ket{\psi^{gW}}\) is admixed with colored orthogonal noise, \(\ket{\psi^{gW}}^{\perp}\), i.e., one can identify conditions on \(J_x, J_y, J_z\) in \(U^{AB}\) which has the ability to activate the dense codeability of \(\rho_{AB}\) and deactivate the quantum advantage of \(\rho_{AE}\).\\ 

\noindent \textbf{Theorem 5.} \textit{ The   unitaries for $\rho_{ABE} = (1-s)\ketbra{\psi^{gW}}{\psi^{gW}}+ s\ketbra{\psi^{gW}}^{\perp}$ satisfy the conditions -- (1) $\mathcal{C}\Big(U^{AB}\Tr_E \Big[\rho_{ABE}\Big](U^{AB})^{\dagger}\Big)$ is independent of $J_z$; and (2) $U^{AB}_*$  is characterized by \(\tan{2 J^*_x} \tan{2 J^*_y} = \frac{(1-2b)+2s(b-\frac{a}{a+b})}{(1-2a-2b)-2s(1-a-b)}\).}\\  

\begin{figure}[t]
    \includegraphics[width=1.0\linewidth]{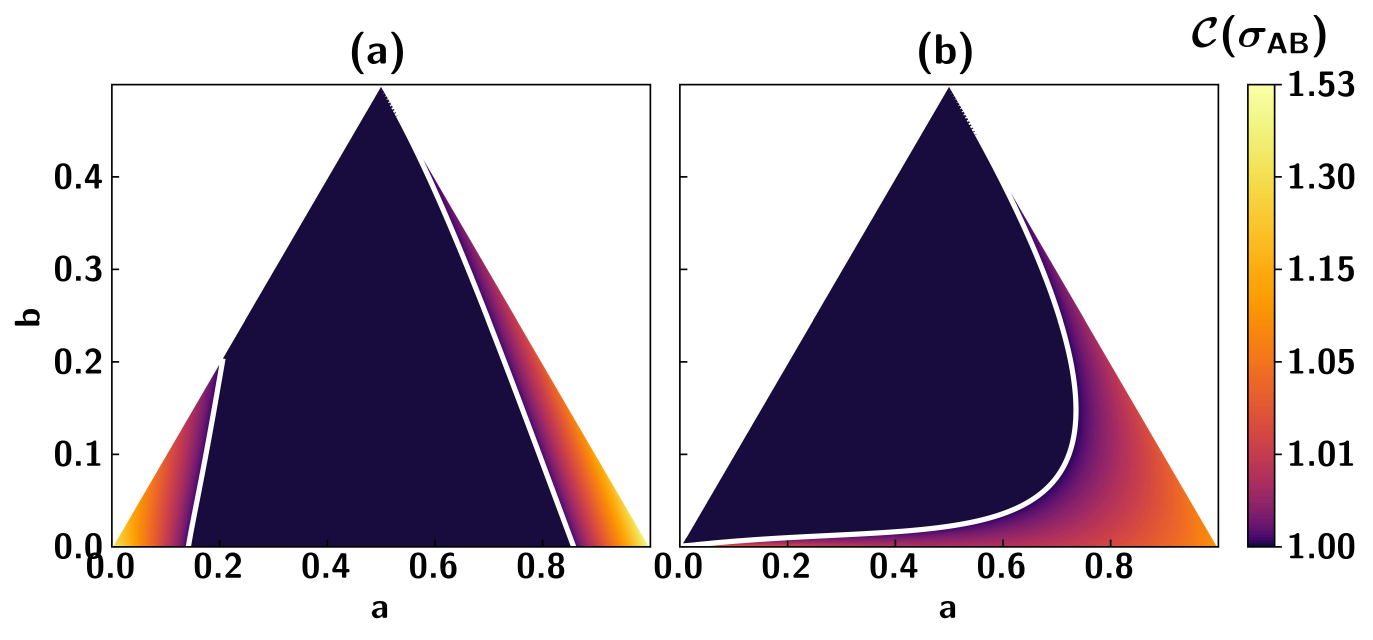}
    \caption{\justifying{  Classification of the parameter space of the orthogonal colored-noise admixed generalized W state according to its capability to generate dense codeability in the reduced state $\sigma_{AB}$ after the application of the optimal unitary. Map plot of \(\mathcal{C}(\sigma_{AB})\) vs  $(a,b)$ for different noise strengths: (a) $s = 0.1$, and (b) $s = 0.8$. Regions with brighter shades denote the region in $(a,b)$ for which the optimal unitary can swap the dense codeability to \(AB\) from \(AE\). The white line (solid) denotes the boundary where the quantum advantage cease to exist. All axes are dimensionless.
    }}
    \label{fig:DC_activationRegion_colornoisy}
\end{figure}

\noindent For proof, see Appendix \ref{A3}. 
In this case, the maximum achievable dense coding capacity between the channel shared between \(A\) and \(B\) is 
\begin{eqnarray}
C_{max} (\sigma_{AB}) =&&2-H(\{0,a-\frac{(a^2-(1-a)b)s}{a+b},{}\nonumber\\&&\frac{1}{2}(1-a(1-s)-\frac{bs}{a+b}-k,\nonumber\\&&\frac{1}{2}(1-a(1-s)-\frac{bs}{a+b}+k)\}),
\label{Eq:Cap_colornoise}
\end{eqnarray}
where
$k=\frac{\sqrt{(a+b-a^2(1-s)-ab(1-s)-bs)^2-4a(1-a-b)(1-s)(a+b)s}}{a+b})$. Here, $\sigma_{AB}$ is the state obtained after application of the swapping unitary $U^{AB}$ on the initial state $\rho_{ABE}$. We also check the region in the state parameter space for which even the optimal unitary cannot activate dense codeability in $AB$ for different fixed values of noise admixing probability (see Fig.   \ref{fig:DC_activationRegion_colornoisy}).\\

\section{Conclusion}
\label{sec:conclusion}

Quantum dense coding (DC) is one of the earliest quantum communication protocols to demonstrate the advantage of entanglement as a resource, enabling the transmission of classical information beyond the classical limit through a shared quantum state. It is also among the first quantum communication protocols to have been experimentally realized across a variety of physical platforms, including photonic~\cite{Mattle_PRL_1996, Shimizu_PRA_1999, Mizuno_PRA_2005, Pan_RMP_2012, Northup_NP_2014, Barreiro_NP_2014, Krenn_PNAS_2016} and atomic systems~\cite{Fang_PRA_2000, Libfried_RMP_2003, Vandersypen_RMP_2005, Yang_JOPB_2007}. In multipartite settings, however, the usefulness of shared entanglement is fundamentally constrained by the dense coding exclusion principle, which states that among channels sharing a common party, at most one can simultaneously exhibit a quantum advantage. Understanding whether such a limitation can nevertheless be exploited in a controlled manner forms the central motivation of this work.\\
 Summarizing, we introduced a multipartite communication set-up, referred to as {\it dense coding swapping}, in which the dense codeability of one bipartite marginal embedded in a multipartite state is transferred to another by applying an appropriate two-qubit unitary operation. For generalized three-qubit pure states, we derived necessary and sufficient conditions for such swapping of dense codeability in terms of the Schmidt coefficients, while for mixed states, we obtained a sufficient criterion based on the correlation matrix. Extending our analysis beyond three-qubits, we demonstrated that multipartite resource states supporting DC swapping are typically asymmetric and possess relatively low genuine multipartite entanglement, which further decreases with increasing system size. Furthermore, for important families of pure and noisy states, we identified the optimal two-qubit quantum gates capable of activating the desired communication channel without involving the compromised party.  Our work thus establishes dense coding swapping as a new operational resource in multipartite quantum communication, particularly for preventing information sharing with dishonest parties. It also illustrates how local two-qubit control can dynamically circumvent the restrictions imposed by the dense coding exclusion principle by redistributing, rather than violating, the available quantum advantage. We anticipate that this framework may stimulate further investigations into the controlled routing of quantum communication resources and their applications in multipartite quantum networks.

\acknowledgements

The authors acknowledge the use of Armadillo~\cite{Sanderson_arma_2019,Sanderson_arma_2025}, and QIClib -- a modern C++ library for general-purpose quantum information processing and quantum computing (\url{https://titaschanda.github.io/QIClib}). A.M. and A.S.D. acknowledge support from the project entitled ``Technology Vertical - Quantum Communication'' under the National Quantum Mission of the Department of Science and Technology (DST)  (Sanction Order No. DST/QTC/NQM/QComm/$2024/2$ (G)). \\
\textbf{Disclaimer: } All the data generated for the analysis are available upon request. Numerical studies are carried out using custom C++ programs developed by the authors and are available upon request.


\bibliography{refer}

@article{Zukowski_PRL_1993,
  title = {``Event-ready-detectors'' Bell experiment via entanglement swapping},
  author = {\ifmmode \dot{Z}\else \.{Z}\fi{}ukowski, M. and Zeilinger, A. and Horne, M. A. and Ekert, A. K.},
  journal = {Phys. Rev. Lett.},
  volume = {71},
  issue = {26},
  pages = {4287--4290},
  numpages = {0},
  year = {1993},
  month = {Dec},
  publisher = {American Physical Society},
  doi = {10.1103/PhysRevLett.71.4287},
  url = {https://link.aps.org/doi/10.1103/PhysRevLett.71.4287}
}

@article{Bose_PRA_1998,
  title = {Multiparticle generalization of entanglement swapping},
  author = {Bose, S. and Vedral, V. and Knight, P. L.},
  journal = {Phys. Rev. A},
  volume = {57},
  issue = {2},
  pages = {822--829},
  numpages = {0},
  year = {1998},
  month = {Feb},
  publisher = {American Physical Society},
  doi = {10.1103/PhysRevA.57.822},
  url = {https://link.aps.org/doi/10.1103/PhysRevA.57.822}
}

@INPROCEEDINGS{Sanderson_arma_2025,
  author={Sanderson, Conrad and Curtin, Ryan},
  booktitle={2025 17th International Conference on Computer and Automation Engineering (ICCAE)}, 
  title={Armadillo: an Efficient Framework for Numerical Linear Algebra}, 
  year={2025},
  volume={},
  number={},
  pages={303-307},
  doi={10.1109/ICCAE64891.2025.10980539}}

@Article{Sanderson_arma_2019,
AUTHOR = {Sanderson, Conrad and Curtin, Ryan},
TITLE = {Practical Sparse Matrices in C++ with Hybrid Storage and Template-Based Expression Optimisation},
JOURNAL = {Mathematical and Computational Applications},
VOLUME = {24},
YEAR = {2019},
NUMBER = {3},
ARTICLE-NUMBER = {70},
URL = {https://www.mdpi.com/2297-8747/24/3/70},
ISSN = {2297-8747},
}

@article{Kendon_PRA_2002,
  title = {Bounds on entanglement in qudit subsystems},
  author = {Kendon, Vivien M. and \ifmmode \dot{Z}\else \.{Z}\fi{}yczkowski, Karol and Munro, William J.},
  journal = {Phys. Rev. A},
  volume = {66},
  issue = {6},
  pages = {062310},
  numpages = {7},
  year = {2002},
  month = {Dec},
  publisher = {American Physical Society},
  doi = {10.1103/PhysRevA.66.062310},
  url = {https://link.aps.org/doi/10.1103/PhysRevA.66.062310}
}

@article{Wei_PRA_2003,
  title = {Geometric measure of entanglement and applications to bipartite and multipartite quantum states},
  author = {Wei, Tzu-Chieh and Goldbart, Paul M.},
  journal = {Phys. Rev. A},
  volume = {68},
  issue = {4},
  pages = {042307},
  numpages = {12},
  year = {2003},
  month = {Oct},
  publisher = {American Physical Society},
  doi = {10.1103/PhysRevA.68.042307},
  url = {https://link.aps.org/doi/10.1103/PhysRevA.68.042307}
}

@article{Nepal_PRA_2013,
  title = {Maximally-dense-coding-capable quantum states},
  author = {Nepal, Rabindra and Prabhu, R. and Sen(De), Aditi and Sen, Ujjwal},
  journal = {Phys. Rev. A},
  volume = {87},
  issue = {3},
  pages = {032336},
  numpages = {6},
  year = {2013},
  month = {Mar},
  publisher = {American Physical Society},
  doi = {10.1103/PhysRevA.87.032336},
  url = {https://link.aps.org/doi/10.1103/PhysRevA.87.032336}
}

@article{Mattle_PRL_1996,
  title = {Dense Coding in Experimental Quantum Communication},
  author = {Mattle, K. and Weinfurter, H. and Kwiat, P. G. and Zeilinger, A.},
  journal = {Phys. Rev. Lett.},
  volume = {76},
  issue = {25},
  pages = {4656--4659},
  numpages = {0},
  year = {1996},
  month = {Jun},
  publisher = {American Physical Society},
  doi = {10.1103/PhysRevLett.76.4656},
  url = {https://link.aps.org/doi/10.1103/PhysRevLett.76.4656}
}

@article{Shimizu_PRA_1999,
  title = {Dense coding in photonic quantum communication with enhanced information capacity},
  author = {Shimizu, K. and Imoto, N. and Mukai, T.},
  journal = {Phys. Rev. A},
  volume = {59},
  issue = {2},
  pages = {1092--1097},
  numpages = {0},
  year = {1999},
  month = {Feb},
  publisher = {American Physical Society},
  doi = {10.1103/PhysRevA.59.1092},
  url = {https://link.aps.org/doi/10.1103/PhysRevA.59.1092}
}

@article{Mizuno_PRA_2005,
  title = {Experimental demonstration of entanglement-assisted coding using a two-mode squeezed vacuum state},
  author = {Mizuno, Jun and Wakui, Kentaro and Furusawa, Akira and Sasaki, Masahide},
  journal = {Phys. Rev. A},
  volume = {71},
  issue = {1},
  pages = {012304},
  numpages = {4},
  year = {2005},
  month = {Jan},
  publisher = {American Physical Society},
  doi = {10.1103/PhysRevA.71.012304},
  url = {https://link.aps.org/doi/10.1103/PhysRevA.71.012304}
}

@article{Pan_RMP_2012,
  title = {Multiphoton entanglement and interferometry},
  author = {Pan, Jian-Wei and Chen, Zeng-Bing and Lu, Chao-Yang and Weinfurter, Harald and Zeilinger, Anton and \ifmmode \dot{Z}\else \.{Z}\fi{}ukowski, Marek},
  journal = {Rev. Mod. Phys.},
  volume = {84},
  issue = {2},
  pages = {777--838},
  numpages = {0},
  year = {2012},
  month = {May},
  publisher = {American Physical Society},
  doi = {10.1103/RevModPhys.84.777},
  url = {https://link.aps.org/doi/10.1103/RevModPhys.84.777}
}

@article{Northup_NP_2014,
  title = {Quantum information transfer using photons},
  author = {Northup, T. E. and  Blatt, R.},
  journal = {Nature Photonics},
  volume = {8},
  issue = {5},
  pages = {356},
  numpages = {8},
  year = {2014},
  month = {May},
  publisher = {Springer Nature},
  doi = {10.1038/nphoton.2014.53},
  url = {https://doi.org/10.1038/nphoton.2014.53}
}

@article{Barreiro_NP_2014,
  title = {Quantum information transfer using photons},
  author = {Barreiro, Julio T. and  Wei, Tzu-Chieh and Kwiat, Paul G.},
  journal = {Nature Physics},
  volume = {4},
  issue = {4},
  pages = {282},
  numpages = {5},
  year = {2014},
  month = {Apr},
  publisher = {Springer Nature},
  doi = {10.1038/nphys919},
  url = {https://doi.org/10.1038/nphys919}
}

@article{Krenn_PNAS_2016,
author = {Krenn, M.  and Handsteiner, J.  and Fink, M.  and Fickler, R.  and Ursin, R.  and Malik, M.  and Zeilinger, A. },
title = {Twisted light transmission over 143 km},
journal = {Proceedings of the National Academy of Sciences},
volume = {113},
number = {48},
pages = {13648-13653},
year = {2016},
doi = {10.1073/pnas.1612023113},
url = {https://www.pnas.org/doi/abs/10.1073/pnas.1612023113}
}

@article{Fang_PRA_2000,
  title = {Experimental implementation of dense coding using nuclear magnetic resonance},
  author = {Fang, X. and Zhu, X. and Feng, M. and Mao, X. and Du, F.},
  journal = {Phys. Rev. A},
  volume = {61},
  issue = {2},
  pages = {022307},
  numpages = {5},
  year = {2000},
  month = {Jan},
  publisher = {American Physical Society},
  doi = {10.1103/PhysRevA.61.022307},
  url = {https://link.aps.org/doi/10.1103/PhysRevA.61.022307}
}

@article{Libfried_RMP_2003,
  title = {Quantum dynamics of single trapped ions},
  author = {Leibfried, D. and Blatt, R. and Monroe, C. and Wineland, D.},
  journal = {Rev. Mod. Phys.},
  volume = {75},
  issue = {1},
  pages = {281--324},
  numpages = {0},
  year = {2003},
  month = {Mar},
  publisher = {American Physical Society},
  doi = {10.1103/RevModPhys.75.281},
  url = {https://link.aps.org/doi/10.1103/RevModPhys.75.281}
}

@article{Vandersypen_RMP_2005,
  title = {NMR techniques for quantum control and computation},
  author = {Vandersypen, L. M. K. and Chuang, I. L.},
  journal = {Rev. Mod. Phys.},
  volume = {76},
  issue = {4},
  pages = {1037--1069},
  numpages = {0},
  year = {2005},
  month = {Jan},
  publisher = {American Physical Society},
  doi = {10.1103/RevModPhys.76.1037},
  url = {https://link.aps.org/doi/10.1103/RevModPhys.76.1037}
}

@article{Yang_JOPB_2007,
doi = {10.1088/0953-4075/40/6/014},
url = {https://dx.doi.org/10.1088/0953-4075/40/6/014},
year = {2007},
month = {mar},
publisher = {},
volume = {40},
number = {6},
pages = {1245},
author = {Yang, W and Gong, Z.},
title = {Practical scheme for quantum dense coding between three parties using microwave radiation in trapped ions},
journal = {Journal of Physics B: Atomic, Molecular and Optical Physics}
}

@article{Aditi_PRA_2003,
  title = {Multiqubit W states lead to stronger nonclassicality than Greenberger-Horne-Zeilinger states},
  author = {Sen(De), Aditi and Sen, Ujjwal and Wie\ifmmode \acute{s}\else \'{s}\fi{}niak, Marcin and Kaszlikowski, Dagomir and \ifmmode \dot{Z}\else \.{Z}\fi{}ukowski, Marek},
  journal = {Phys. Rev. A},
  volume = {68},
  issue = {6},
  pages = {062306},
  numpages = {7},
  year = {2003},
  month = {Dec},
  publisher = {American Physical Society},
  doi = {10.1103/PhysRevA.68.062306},
  url = {https://link.aps.org/doi/10.1103/PhysRevA.68.062306}
}

@article{Greenberger_arXiv_2007,
  doi = {10.48550/arXiv.0712.0921},
  
  url = {https://arxiv.org/abs/0712.0921},
  
  author = {Greenberger, D. M. and Horne, M. A. and Zeilinger, A.},
  
   title = {Going Beyond Bell's Theorem},
  
  publisher = {arXiv},
  
  year = {2007},
  
  copyright = {Assumed arXiv.org perpetual, non-exclusive license to distribute this article for submissions made before January 2004}
}

@incollection {Lomonaco_arXiv_2004,
    AUTHOR = {Yimsiriwattana, A. and Lomonaco Jr., S. J.},
     TITLE = {Generalized {GHZ} states and distributed quantum computing},
 BOOKTITLE = {Coding theory and quantum computing},
    SERIES = {Contemp. Math.},
    VOLUME = {381},
     PAGES = {131--147},
 PUBLISHER = {Amer. Math. Soc., Providence, RI},
      YEAR = {2005},
      ISBN = {0-8218-3600-5},
       DOI = {10.1090/conm/381/07096},
       URL = {https://doi.org/10.1090/conm/381/07096},
}

@article{Chiuri_PRL_2012,
  title = {Experimental Quantum Networking Protocols via Four-Qubit Hyperentangled Dicke States},
  author = {Chiuri, A. and Greganti, C. and Paternostro, M. and Vallone, G. and Mataloni, P.},
  journal = {Phys. Rev. Lett.},
  volume = {109},
  issue = {17},
  pages = {173604},
  numpages = {5},
  year = {2012},
  month = {Oct},
  publisher = {American Physical Society},
  doi = {10.1103/PhysRevLett.109.173604},
  url = {https://link.aps.org/doi/10.1103/PhysRevLett.109.173604}
}

@article{Bernd_PRL_2014,
  title = {Detecting Multiparticle Entanglement of Dicke States},
  author = {L\"ucke, Bernd and Peise, Jan and Vitagliano, Giuseppe and Arlt, Jan and Santos, Luis and T\'oth, G\'eza and Klempt, Carsten},
  journal = {Phys. Rev. Lett.},
  volume = {112},
  issue = {15},
  pages = {155304},
  numpages = {5},
  year = {2014},
  month = {Apr},
  publisher = {American Physical Society},
  doi = {10.1103/PhysRevLett.112.155304},
  url = {https://link.aps.org/doi/10.1103/PhysRevLett.112.155304}
}

@article{Bergmann_2013,
doi = {10.1088/1751-8113/46/38/385304},
url = {https://doi.org/10.1088/1751-8113/46/38/385304},
year = {2013},
month = {sep},
publisher = {IOP Publishing},
volume = {46},
number = {38},
pages = {385304},
author = {Bergmann, Marcel and Gühne, Otfried},
title = {Entanglement criteria for Dicke states},
journal = {J. Phys. A: Math. Theor.},
}

@article{Dicke,
  title = {Coherence in Spontaneous Radiation Processes},
  author = {Dicke, R. H.},
  journal = {Phys. Rev.},
  volume = {93},
  issue = {1},
  pages = {99--110},
  numpages = {0},
  year = {1954},
  month = {Jan},
  publisher = {American Physical Society},
  doi = {10.1103/PhysRev.93.99},
  url = {https://link.aps.org/doi/10.1103/PhysRev.93.99}
}

@article{Asutosh_PLA_2017,
title = {Forbidden regimes in the distribution of bipartite quantum correlations due to multiparty entanglement},
journal = {Phys. Lett. A},
volume = {381},
number = {20},
pages = {1701-1709},
year = {2017},
issn = {0375-9601},
doi = {https://doi.org/10.1016/j.physleta.2017.03.026},
url = {https://www.sciencedirect.com/science/article/pii/S0375960116316280},
author = {Asutosh Kumar and Himadri Shekhar Dhar and R. Prabhu and Aditi Sen(De) and Ujjwal Sen},
}

@article{Audenart_JMP_2005,
    author = {Audenaert, Koenraad M. R. and Eisert, Jens},
    title = {Continuity bounds on the quantum relative entropy},
    journal = {J. Math. Phys.},
    volume = {46},
    number = {10},
    pages = {102104},
    year = {2005},
    month = {10},
    issn = {0022-2488},
    doi = {10.1063/1.2044667},
    url = {https://doi.org/10.1063/1.2044667},
}

@article{Dur_PRA_2000,
  title = {Three qubits can be entangled in two inequivalent ways},
  author = {D\"ur, W. and Vidal, G. and Cirac, J. I.},
  journal = {Phys. Rev. A},
  volume = {62},
  issue = {6},
  pages = {062314},
  numpages = {12},
  year = {2000},
  month = {Nov},
  publisher = {American Physical Society},
  doi = {10.1103/PhysRevA.62.062314},
  url = {https://link.aps.org/doi/10.1103/PhysRevA.62.062314}
}

@article{Vatan_PRA_2004,
  title = {Optimal quantum circuits for general two-qubit gates},
  author = {Vatan, Farrokh and Williams, Colin},
  journal = {Phys. Rev. A},
  volume = {69},
  issue = {3},
  pages = {032315},
  numpages = {5},
  year = {2004},
  month = {Mar},
  publisher = {American Physical Society},
  doi = {10.1103/PhysRevA.69.032315},
  url = {https://link.aps.org/doi/10.1103/PhysRevA.69.032315}
}

@article{Ratul_PRA_2020,
  title = {Uniform decoherence effect on localizable entanglement in random multiqubit pure states},
  author = {Banerjee, Ratul and Pal, Amit Kumar and Sen(De), Aditi},
  journal = {Phys. Rev. A},
  volume = {101},
  issue = {4},
  pages = {042339},
  numpages = {12},
  year = {2020},
  month = {Apr},
  publisher = {American Physical Society},
  doi = {10.1103/PhysRevA.101.042339},
  url = {https://link.aps.org/doi/10.1103/PhysRevA.101.042339}
}

@article{Tamoghna_PRA_2016,
  title = {Generalized geometric measure of entanglement for multiparty mixed states},
  author = {Das, Tamoghna and Roy, Sudipto Singha and Bagchi, Shrobona and Misra, Avijit and Sen(De), Aditi and Sen, Ujjwal},
  journal = {Phys. Rev. A},
  volume = {94},
  issue = {2},
  pages = {022336},
  numpages = {8},
  year = {2016},
  month = {Aug},
  publisher = {American Physical Society},
  doi = {10.1103/PhysRevA.94.022336},
  url = {https://link.aps.org/doi/10.1103/PhysRevA.94.022336}
}

@article{Aditiggm_PRA_2010,
  title = {Channel capacities versus entanglement measures in multiparty quantum states},
  author = {Sen(De), Aditi and Sen, Ujjwal},
  journal = {Phys. Rev. A},
  volume = {81},
  issue = {1},
  pages = {012308},
  numpages = {6},
  year = {2010},
  month = {Jan},
  publisher = {American Physical Society},
  doi = {10.1103/PhysRevA.81.012308},
  url = {https://link.aps.org/doi/10.1103/PhysRevA.81.012308}
}

@article{Page_PRL_1993,
  title = {Average entropy of a subsystem},
  author = {Page, Don N.},
  journal = {Phys. Rev. Lett.},
  volume = {71},
  issue = {9},
  pages = {1291--1294},
  numpages = {0},
  year = {1993},
  month = {Aug},
  publisher = {American Physical Society},
  doi = {10.1103/PhysRevLett.71.1291},
  url = {https://link.aps.org/doi/10.1103/PhysRevLett.71.1291}
}

@article{Zhang_PRA_2003,
  title = {Geometric theory of nonlocal two-qubit operations},
  author = {Zhang, Jun and Vala, Jiri and Sastry, Shankar and Whaley, K. Birgitta},
  journal = {Phys. Rev. A},
  volume = {67},
  issue = {4},
  pages = {042313},
  numpages = {18},
  year = {2003},
  month = {Apr},
  publisher = {American Physical Society},
  doi = {10.1103/PhysRevA.67.042313},
  url = {https://link.aps.org/doi/10.1103/PhysRevA.67.042313}
}

@article{Hausladen_PRA_1996,
  title = {Classical information capacity of a quantum channel},
  author = {Hausladen, Paul and Jozsa, Richard and Schumacher, Benjamin and Westmoreland, Michael and Wootters, William K.},
  journal = {Phys. Rev. A},
  volume = {54},
  issue = {3},
  pages = {1869--1876},
  numpages = {0},
  year = {1996},
  month = {Sep},
  publisher = {American Physical Society},
  doi = {10.1103/PhysRevA.54.1869},
  url = {https://link.aps.org/doi/10.1103/PhysRevA.54.1869}
}

@article{Bowen_PRA_2001,
  title = {Classical information capacity of superdense coding},
  author = {Bowen, G.},
  journal = {Phys. Rev. A},
  volume = {63},
  issue = {2},
  pages = {022302},
  numpages = {4},
  year = {2001},
  month = {Jan},
  publisher = {American Physical Society},
  doi = {10.1103/PhysRevA.63.022302},
  url = {https://link.aps.org/doi/10.1103/PhysRevA.63.022302}
}

@article{Horodecki_arxiv_2001,
author = {Horodecki, M. and Horodecki, P. and Horodecki, R. and Leung, D. W. and Terhal, B. M.},
title = {Classical capacity of a noiseless quantum channel assisted by noisy entanglement},
year = {2001},
volume = {1},
number = {3},
journal = {Quantum Info. Comput.},
month = {oct},
pages = {70–78},
numpages = {9},
doi = {https://arxiv.org/abs/quant-ph/0106080}
}

@article{horodecki2009quantum,
  title={Quantum entanglement},
  author={Horodecki, Ryszard and Horodecki, Pawe{\l} and Horodecki, Micha{\l} and Horodecki, Karol},
  journal={Reviews of modern physics},
  volume={81},
  number={2},
  pages={865},
  year={2009},
  doi = {https://doi.org/10.1103/RevModPhys.81.865},
  publisher={APS}
}

@article{bennett1992communication,
  title={Communication via one-and two-particle operators on Einstein-Podolsky-Rosen states},
  author={Bennett, Charles H and Wiesner, Stephen J},
  journal={Physical Review Letters},
  volume={69},
  number={20},
  pages={2881},
  year={1992},
  doi = {https://doi.org/10.1103/PhysRevLett.69.2881},
  publisher={APS}
}

@article{srivastava2019one,
  title={One-shot conclusive multiport quantum dense coding capacities},
  author={Srivastava, Chirag and Bera, Anindita and Sen, Aditi and Sen, Ujjwal and others},
  journal={Physical Review A},
  volume={100},
  number={5},
  pages={052304},
  year={2019},
  doi = {https://doi.org/10.1103/PhysRevA.100.052304},
  publisher={APS}
}

@article{vempati2021witnessing,
  title={Witnessing negative conditional entropy},
  author={Vempati, Mahathi and Ganguly, Nirman and Chakrabarty, Indranil and Pati, Arun K},
  journal={Physical Review A},
  volume={104},
  number={1},
  pages={012417},
  year={2021},
  doi = {https://doi.org/10.1103/PhysRevA.104.012417},
  publisher={APS}
}

@article{patro2017non,
  title={Non-negativity of conditional von Neumann entropy and global unitary operations},
  author={Patro, Subhasree and Chakrabarty, Indranil and Ganguly, Nirman},
  journal={Physical Review A},
  volume={96},
  number={6},
  pages={062102},
  year={2017},
  doi = {https://doi.org/10.1103/PhysRevA.96.062102},
  publisher={APS}
}

@article{srinidhi2024quantum,
  title={Quantum channels that destroy negative conditional entropy},
  author={Srinidhi, PV and Chakrabarty, Indranil and Bhattacharya, Samyadeb and Ganguly, Nirman},
  journal={Physical Review A},
  volume={110},
  number={4},
  pages={042423},
  year={2024},
  doi = {https://doi.org/10.1103/PhysRevA.110.042423},
  publisher={APS}
}

@article{moravvcikova2010entanglement,
  title={Entanglement-annihilating and entanglement-breaking channels},
  author={Morav{\v{c}}{\'\i}kov{\'a}, Lenka and Ziman, M{\'a}rio},
  journal={Journal of Physics A: Mathematical and Theoretical},
  volume={43},
  number={27},
  pages={275306},
  doi = {https://iopscience.iop.org/article/10.1088/1751-8113/43/27/275306/pdf},
  year={2010}
}

@article{vedral1997quantifying,
  title={Quantifying entanglement},
  author={Vedral, Vlatko and Plenio, Martin B and Rippin, Michael A and Knight, Peter L},
  journal={Physical Review Letters},
  volume={78},
  number={12},
  pages={2275},
  year={1997},
  doi = {https://doi.org/10.1103/PhysRevLett.78.2275},
  publisher={APS}
}

@article{prabhu2012exclusion,
  title = {Exclusion principle for quantum dense coding},
  author = {Prabhu, R. and Pati, Arun Kumar and Sen(De), Aditi and Sen, Ujjwal},
  journal = {Phys. Rev. A},
  volume = {87},
  issue = {5},
  pages = {052319},
  numpages = {6},
  year = {2013},
  month = {May},
  publisher = {American Physical Society},
  doi = {10.1103/PhysRevA.87.052319},
  url = {https://link.aps.org/doi/10.1103/PhysRevA.87.052319}
}

@article{cerf1997negative,
  title={Negative entropy and information in quantum mechanics},
  author={Cerf, Nicolas J and Adami, Chris},
  journal={Physical Review Letters},
  volume={79},
  number={26},
  pages={5194},
  year={1997},
  doi = {https://doi.org/10.1103/PhysRevLett.79.5194},
  publisher={APS}
}

@article{horodecki2005partial,
  title={Partial quantum information},
  author={Horodecki, Micha{\l} and Oppenheim, Jonathan and Winter, Andreas},
  journal={Nature},
  volume={436},
  number={7051},
  pages={673--676},
  year={2005},
  doi = {https://doi.org/10.1038/nature03909},
  publisher={Nature Publishing Group UK London}
}

@article{bruss93sen,
  title={Distributed Quantum Dense Coding},
  author={Bru{\ss}, D and D’Ariano, GM and Lewenstein, M and Macchiavello, C and Sen(De), A. and Sen, U.},
  journal={Phys. Rev. Lett},
  volume={93},
  doi = {https://doi.org/10.1103/PhysRevLett.93.210501},
  pages={210501}
}

@article{devetak2005distillation,
  title={Distillation of secret key and entanglement from quantum states},
  author={Devetak, Igor and Winter, Andreas},
  journal={Proceedings of the Royal Society A: Mathematical, Physical and engineering sciences},
  volume={461},
  number={2053},
  pages={207--235},
  year={2005},
  doi = {https://doi.org/10.1098/rspa.2004.1372},
  publisher={The Royal Society}
}

@article{gyongyosi2012properties,
  title={Properties of the quantum channel},
  author={Gyongyosi, Laszlo and Imre, Sandor},
  journal={arXiv preprint arXiv:1208.1270},
  doi = {https://doi.org/10.48550/arXiv.1208.1270},
  year={2012}
}

@article{hao2001controlled,
  title={Controlled dense coding using the Greenberger-Horne-Zeilinger state},
  author={Hao, Jiu-Cang and Li, Chuan-Feng and Guo, Guang-Can},
  journal={Physical Review A},
  volume={63},
  number={5},
  pages={054301},
  year={2001},
  publisher={APS}
}

@article{zhang2002controlled,
  title={Controlled dense coding for continuous variables using three-particle entangled states},
  author={Zhang, Jing and Xie, Changde and Peng, Kunchi},
  journal={Physical Review A},
  volume={66},
  number={3},
  pages={032318},
  year={2002},
  publisher={APS}
}

@article{jing2003experimental,
  title={Experimental Demonstration of Tripartite Entanglement and Controlled Dense Coding<? format?> for Continuous Variables},
  author={Jing, Jietai and Zhang, Jing and Yan, Ying and Zhao, Fagang and Xie, Changde and Peng, Kunchi},
  journal={Physical review letters},
  volume={90},
  number={16},
  pages={167903},
  year={2003},
  publisher={APS}
}

@article{fu2006controlled,
  title={Controlled quantum dense coding in a four-particle non-maximally entangled state via local measurements},
  author={Fu, Chang-Bao and Xia, Yan and Liu, Bo-Xue and Zhang, Shou and Yeon, Kyu-Hwang and Um, Chung-In},
  journal={arXiv preprint quant-ph/0601144},
  year={2006}
}

@article{oh2017minimal,
  title={Minimal control power of controlled dense coding and genuine tripartite entanglement},
  author={Oh, Changhun and Kim, Hoyong and Jeong, Kabgyun and Jeong, Hyunseok},
  journal={Scientific Reports},
  volume={7},
  number={1},
  pages={3765},
  year={2017},
  publisher={Nature Publishing Group UK London}
}

@article{pati2005probabilistic,
  title = {Probabilistic superdense coding},
  author = {Pati, A. K. and Parashar, P. and Agrawal, P.},
  journal = {Phys. Rev. A},
  volume = {72},
  issue = {1},
  pages = {012329},
  numpages = {6},
  year = {2005},
  month = {Jul},
  publisher = {American Physical Society},
  doi = {10.1103/PhysRevA.72.012329},
  url = {https://link.aps.org/doi/10.1103/PhysRevA.72.012329}
}

@article{zyczkowski1998volume,
  title={Volume of the set of separable states},
  author={{\.Z}yczkowski, Karol and Horodecki, Pawe{\l} and Sanpera, Anna and Lewenstein, Maciej},
  journal={Physical Review A},
  volume={58},
  number={2},
  pages={883},
  year={1998},
  publisher={APS}
}

@article{berta2010uncertainty,
  title={The uncertainty principle in the presence of quantum memory},
  author={Berta, Mario and Christandl, Matthias and Colbeck, Roger and Renes, Joseph M and Renner, Renato},
  journal={Nature Physics},
  volume={6},
  number={9},
  pages={659--662},
  year={2010},
  doi = {https://doi.org/10.1038/nphys1734},
  publisher={Nature Publishing Group UK London}
}

@article{azuma2018black,
  title={Do black holes store negative entropy?},
  author={Azuma, Koji and Subramanian, Sathyawageeswar and Kato, Go},
  journal={arXiv preprint arXiv:1807.06753},
  doi = {https://doi.org/10.48550/arXiv.1807.06753},
  year={2018}
}

@article{azuma2020second,
  title={Second law of black hole thermodynamics},
  author={Azuma, Koji and Kato, Go},
  journal={arXiv preprint arXiv:2001.02897},
  doi = {https://doi.org/10.48550/arXiv.2001.02897},
  year={2020}
}

@article{rio2011thermodynamic,
  title={The thermodynamic meaning of negative entropy},
  author={Rio, L{\'\i}dia del and {\AA}berg, Johan and Renner, Renato and Dahlsten, Oscar and Vedral, Vlatko},
  journal={Nature},
  volume={474},
  number={7349},
  pages={61--63},
  year={2011},
  doi = {https://doi.org/10.1038/nature10123},
  publisher={Nature Publishing Group UK London}
}

@article{cerf1999quantum,
  title={Quantum extension of conditional probability},
  author={Cerf, Nicolas J and Adami, Christoph},
  journal={Physical Review A},
  volume={60},
  number={2},
  pages={893},
  year={1999},
  doi = {https://doi.org/10.1103/PhysRevA.60.893},
  publisher={APS}
}

@article{friis2017geometry,
  title={Geometry of two-qubit states with negative conditional entropy},
  author={Friis, Nicolai and Bulusu, Sridhar and Bertlmann, Reinhold A},
  journal={Journal of Physics A: Mathematical and Theoretical},
  volume={50},
  number={12},
  pages={125301},
  year={2017},
  url = {https://iopscience.iop.org/article/10.1088/1751-8121/aa5dfd/pdf},
  publisher={IOP Publishing}
}

@article{bruss2006dense,
  title={Dense coding with multipartite quantum states},
  author={Bru{\ss}, Dagmar and Lewenstein, Maciej and Sen, Aditi and Sen, Ujjwal and D'ARIANO, GIACOMO MAURO and Macchiavello, Chiara},
  journal={International Journal of Quantum Information},
  volume={4},
  number={03},
  pages={415--428},
  year={2006},
  doi = {https://doi.org/10.1142/S0219749906001888},
  publisher={World Scientific}
}

@article{nepal2013maximally,
  title={Maximally-dense-coding-capable quantum states},
  author={Nepal, Rabindra and Prabhu, R and Sen, Aditi and Sen, Ujjwal},
  journal={Physical Review A—Atomic, Molecular, and Optical Physics},
  volume={87},
  number={3},
  pages={032336},
  year={2013},
  doi = {https://doi.org/10.1103/PhysRevA.87.032336},
  publisher={APS}
}

@article{das2014multipartite,
  title={Multipartite dense coding versus quantum correlation: Noise inverts relative capability of information transfer},
  author={Das, Tamoghna and Prabhu, R and Sen, Aditi and Sen, Ujjwal},
  journal={Physical Review A},
  volume={90},
  number={2},
  pages={022319},
  year={2014},
  doi = {https://doi.org/10.1103/PhysRevA.90.022319},
  publisher={APS}
}

@article{Shadman2013,
url = {https://doi.org/10.2478/qmetro-2013-0004},
title = {A review on super dense coding over covariant noisy
channels},
author = {Z. Shadman and H. Kampermann and C. Macchiavello and D. Bruß},
pages = {21--33},
volume = {1},
number = {1},
journal = {Quantum Measurements and Quantum Metrology},
doi = {doi:10.2478/qmetro-2013-0004},
year = {2013}
}

@article{Shadman2011,
  title = {Optimal superdense coding over memory channels},
  author = {Shadman, Z. and Kampermann, H. and Bru\ss{}, D. and Macchiavello, C.},
  journal = {Phys. Rev. A},
  volume = {84},
  issue = {4},
  pages = {042309},
  numpages = {8},
  year = {2011},
  month = {Oct},
  publisher = {American Physical Society},
  doi = {10.1103/PhysRevA.84.042309},
  url = {https://link.aps.org/doi/10.1103/PhysRevA.84.042309}
}

@article{Shadman_NJP_2010,
doi = {10.1088/1367-2630/12/7/073042},
url = {https://dx.doi.org/10.1088/1367-2630/12/7/073042},
year = {2010},
month = {jul},
publisher = {},
volume = {12},
number = {7},
pages = {073042},
author = {Shadman, Z. and Kampermann, H. and Macchiavello, C. and Bruß, D.},
title = {Optimal super dense coding over noisy quantum channels},
journal = {New Journal of Physics}
}

@article{Shadman_PRA_2012,
  title = {Distributed superdense coding over noisy channels},
  author = {Shadman, Z. and Kampermann, H. and Bru\ss{}, D. and Macchiavello, C.},
  journal = {Phys. Rev. A},
  volume = {85},
  issue = {5},
  pages = {052306},
  numpages = {9},
  year = {2012},
  month = {May},
  publisher = {American Physical Society},
  doi = {10.1103/PhysRevA.85.052306},
  url = {https://link.aps.org/doi/10.1103/PhysRevA.85.052306}
}

@article{samanta2024continuous,
  title={Continuous variable dense coding under realistic non-ideal scenarios},
  author={Samanta, Mrinmoy and Patra, Ayan and Gupta, Rivu and De, Aditi Sen},
  journal={arXiv preprint arXiv:2407.07609},
  doi = {https://doi.org/10.48550/arXiv.2407.07609},
  year={2024}
}

@article{acin2001classification,
  title={Classification of mixed three-qubit states},
  author={Ac{\'\i}n, Antonio and Bru{\ss}, Dagmar and Lewenstein, Maciej and Sanpera, Anna},
  journal={Physical Review Letters},
  volume={87},
  number={4},
  pages={040401},
  year={2001},
  doi = {https://doi.org/10.1103/PhysRevLett.87.040401},
  publisher={APS}
}

@article{das2015distributed,
  title={Distributed quantum dense coding with two receivers in noisy environments},
  author={Das, Tamoghna and Prabhu, R and Sen, Aditi and Sen, Ujjwal},
  journal={Physical Review A},
  volume={92},
  number={5},
  pages={052330},
  year={2015},
  doi = {https://doi.org/10.1103/PhysRevA.92.052330},
  publisher={APS}
}

@article{horodecki2007quantum,
  title={Quantum state merging and negative information},
  author={Horodecki, Micha{\l} and Oppenheim, Jonathan and Winter, Andreas},
  journal={Communications in Mathematical Physics},
  volume={269},
  number={1},
  pages={107--136},
  year={2007},
  doi = {https://doi.org/10.1007/s00220-006-0118-x},
  publisher={Springer}
}

@article{muhuri2023information,
  title={Information theoretic resource-breaking channels},
  author={Muhuri, Abhishek and Patra, Ayan and Gupta, Rivu and De, Aditi Sen},
  journal={arXiv preprint arXiv:2309.03108},
  doi = {https://doi.org/10.48550/arXiv.2309.03108},
  year={2023}
}

@article{hirota2020pinskerinequality,
      title={Application of quantum Pinsker inequality to quantum communications}, 
      author={Osamu Hirota},
      journal={arXiv preprint arXiv:2005.04553},
      doi={https://arxiv.org/abs/2005.04553}, 
      year={2020}
}

@article{vempati2022unital,
  title={A-unital operations and quantum conditional entropy},
  author={Vempati, Mahathi and Shah, Saumya and Ganguly, Nirman and Chakrabarty, Indranil},
  journal={Quantum},
  volume={6},
  pages={641},
  year={2022},
  doi = {https://doi.org/10.22331/q-2022-02-02-641},
  publisher={Verein zur F{\"o}rderung des Open Access Publizierens in den Quantenwissenschaften}
}

@article{roy2018deterministic,
  title={Deterministic quantum dense coding networks},
  author={Roy, Saptarshi and Chanda, Titas and Das, Tamoghna and De, Aditi Sen and Sen, Ujjwal},
  journal={Physics Letters A},
  volume={382},
  number={26},
  pages={1709--1715},
  year={2018},
  doi = {https://doi.org/10.1016/j.physleta.2018.04.033},
  publisher={Elsevier}
}

@article{horodecki2003entanglement,
  title={Entanglement breaking channels},
  author={Horodecki, Michael and Shor, Peter W and Ruskai, Mary Beth},
  journal={Reviews in Mathematical Physics},
  volume={15},
  number={06},
  pages={629--641},
  year={2003},
  doi = {https://doi.org/10.1142/S0129055X03001709},
  publisher={World Scientific}
}

\appendix
\begin{widetext}

\section{Schmidt form of generalized W states}\label{A}

One of the exemplary multipartite pure states is the generalized W states, which has the following form in computational basis,
\begin{eqnarray}
    |gW\rangle^{N+1} = \sum_i b_i \mathcal{P} [|0\rangle^{\otimes N} |1\rangle]
\end{eqnarray}
where, $b_i$-s are chosen to be real and $\mathcal{P}$ denotes the permutation operator which permutes the vector $|1\rangle$ in different positions. In tripartite scenario, it reduces to $\ket{\psi^{gW}} = \sqrt{a}\ket{001}+\sqrt{b}\ket{010}+\sqrt{1-a-b}\ket{100}$, where $a\geq 0$, $b\geq 0$ and $1-a-b \geq 0$. To find its Schmidt form as given in Eq.(~\ref{Equation1}), we find its Schmidt bases in the following steps:
\begin{eqnarray}
    \ket{\psi^{gW}} &=& \sqrt{a}\ket{001}+\sqrt{b}\ket{010}+\sqrt{1-a-b}\ket{100}\nonumber\\
    &=& \ket{0}\otimes (\sqrt{a}\ket{01}+\sqrt{b}\ket{10}) + \ket{1}\otimes (\sqrt{1-a-b}\ket{00})
\end{eqnarray}
We introduce the coefficient matrices $T_0$ and $T_1$ such that $(T_i)_{jk} = t_{ijk}$ where $t_{ijk}$ is the coefficient before the vector $\ket{ijk}$, i.e., in our case,
\begin{eqnarray}
    T_0 = 
    \begin{pmatrix}
        0 & \sqrt{a}\\
        \sqrt{b} & 0
    \end{pmatrix}
\end{eqnarray}
\begin{eqnarray}
    T_1 = 
    \begin{pmatrix}
        \sqrt{1-a-b} & 0\\
        0 & 0
    \end{pmatrix}
\end{eqnarray}
We now consider the unitary transformation $U_A=\begin{bmatrix} u_{00} & u_{01} \\ u_{10} & u_{11} \end{bmatrix}$ on the qubit \(A\) which transforms: $T_0' = u_{00} T_0 + u_{01} T_1$ and $T_1' = u_{10} T_0 + u_{11} T_1$
\begin{eqnarray}
     T_0' = 
    \begin{pmatrix}
        u_{01}\sqrt{1-a-b} & u_{00}\sqrt{a}\\
        u_{00}\sqrt{b} & 0
    \end{pmatrix}
\end{eqnarray}
To find the unitary that reduces the state to Schmidt form, which has only relevant parameters, we set $\det(T_0') = 0$, which implies $u_{00} = 0$. Thus the unitary on \(A\) has the form $U_A=\begin{bmatrix} 0 & u_{01} \\ u_{10} & u_{11} \end{bmatrix}$. Putting the condition of unitary:
\begin{eqnarray}
    U_A^{\dagger}U_A = 
    \begin{pmatrix}
        |u_{10}|^2 & u_{10}^*u_{11} \\
        u_{10}u_{11}^* & |u_{01}|^2 + |u_{11}|^2
    \end{pmatrix}
    =
    \begin{pmatrix}
        1 & 0\\
        0 & 1 
    \end{pmatrix}
\end{eqnarray}
 also, 
 \begin{eqnarray}
    U_AU_A^{\dagger} = 
    \begin{pmatrix}
        |u_{01}|^2 & u_{01}u_{11}^* \\
        u_{01}^*u_{11} & |u_{10}|^2 + |u_{11}|^2
    \end{pmatrix}
    =
    \begin{pmatrix}
        1 & 0\\
        0 & 1 
    \end{pmatrix}
\end{eqnarray}
which is possible only when $u_{11}=0$ and $|u_{01}|^2 = |u_{10}|^2 = 1$. This allows us to consider that the unitary $U_A = \begin{bmatrix} 0 & 1 \\ 1 & 0 \end{bmatrix}$. Hence, $T_0' = \begin{bmatrix} \sqrt{1-a-b} & 0 \\ 0 & 0 \end{bmatrix}$ implying $\lambda_0 = \sqrt{1-a-b}$. Correspondingly, $T_1' = \begin{bmatrix} 0 & \sqrt{a} \\ \sqrt{b} & 0\end{bmatrix}$. Thus, the state $\ket{\psi^{gW}}$ in can be cast into the Schmidt form in the following manner:
\begin{eqnarray}
    \ket{\psi^{gW}} &=& \sqrt{a}\ket{001}+\sqrt{b}\ket{010}+\sqrt{1-a-b}\ket{100}\nonumber\\
    &=& (U_A\otimes U_B \otimes U_C) (\sqrt{1-a-b}\ket{000}+\sqrt{a}\ket{101}+\sqrt{b}\ket{110})
    \label{Eq:gW_AcinForm}
\end{eqnarray}
where, $U_A = \sigma_x = \begin{bmatrix} 0 & 1 \\ 1 & 0 \end{bmatrix}$, while $U_B = U_C = \mathbb{I}_2$. This local unitary transformation allows us to determine the Schmidt coefficients of the tripartite generalized W state: $\lambda_0 = \sqrt{1-a-b}$, $\lambda_2=\sqrt{a}$ and $\lambda_3 = \sqrt{b}$.

\section{Calculations of part $2$ of Theorem $3$}\label{A1}
Suppose $J^*_x$ and $J^*_y$ are the optimal unitary parameter for which maximum amount of dense codeability can be activated in $\rho^{gW}_{AB}$.
Equating the eigenvalues of $\sigma^{gW}_B$, we obtain,
\begin{eqnarray}
    &&a\cos^2(J^*_x-J^*_y)+b\sin^2(J^*_x+J^*_y)+(1-a-b)\cos^2(J^*_x+J^*_y)\nonumber\\
    &=&a\sin^2(J^*_x-J^*_y)+b\cos^2(J^*_x+J^*_y)+(1-a-b)\sin^2(J^*_x+J^*_y)\nonumber\\
    &\implies& a\cos{2(J^*_x-J^*_y)+(1-a-2b)\cos{2(J^*_x+J^*_y)} = 0}\nonumber\\
    &\implies& (1-2b)\cos{2J^*_x}\cos{2J^*_y} - (1-2a-2b)\sin{2J^*_x}\sin{2J^*_y} = 0\nonumber\\
    &\implies& \tan{2 J^*_x} \tan{2 J^*_y} = \frac{1-2b}{1-2a-2b}
\end{eqnarray}

\section{Proof of Theorem $4$}\label{A2}

The proof follows the same line of argument as in Theorem 3: $1.$ The mixed state $\sigma_{AB} = \Tr_E\Big[(U^{AB}\otimes\mathbb{I}_2^{E})\rho_{ABE}(U^{AB}\otimes\mathbb{I}_2^{E})^{\dagger}\Big]$ resulting from tracing out the subsystem '$E$' does not involve any $J_z$:

\begin{eqnarray}
    &&\sigma_{AB} = \Tr_E\Big[(U^{AB}\otimes\mathbb{I}_2^{E})\rho_{ABE}(U^{AB}\otimes\mathbb{I}_2^{E})^{\dagger}\Big]\nonumber\\
    &=& U^{AB} \rho_{AB} (U^{AB})^{\dagger}\nonumber\\
    &=&
    \begin{bmatrix}
        A_1+A_2 & 0 & 0 & B\\
        0 & C_1-C_2 & D & 0 \\
        0 & D^* &C_1+C_2 & 0\\
        B^* & 0 & 0 &  A_1-A_2
    \end{bmatrix}\nonumber\\
    \label{Eq:UonAB_mix}
\end{eqnarray}
where $A_1 = \frac{1}{4}(2a(1-s)+s)$, $A_2 = \frac{1}{4}2a(1-s)\cos(2J_{x-y})$, $B=\frac{i}{2} a \sin(2J_{x-y})$, $C_1 = \frac{1}{8}(4-4a(1-s)-2s)$, $C_2=\frac{1}{8}(4(1-a-2b)(1-s)\cos(2J_{x+y}))$ and $D = \frac{1}{2}((1-s)(2\sqrt{b(1-a-b)}-i (1-a-b)\sin(2J_{x+y}))$, with $J_{x\pm y}=J_x \pm J_y$ which is independent of the parameter $J_z$, implying its capacity does not depend on $J_z$. $\blacksquare$\\
    
\noindent $2.$ The reduced state $\sigma_B = \Tr_A[U^{AB}\rho_{AB}(U^{AB})^{\dagger}]$ is diagonal in computational basis and its eigenvalues are: $\{\frac{1}{2}(1-a(1-s)\cos2J_{x-y}-(1-a-2b)(1-s)\cos2J_{x+y}), \frac{1}{2}(1+a(1-s)\cos2J_{x-y}+(1-a-2b)(1-s)\cos2J_{x+y})\}$. Important thing to notice is that the entropy of $\rho_{AB}$ does not change under the swapping unitary, only the spectrum of the reduced state of party $B$ changes, and thus decides the optimality of the unitary. The eigenvalues of $\sigma_B$ become equal at Eq.(~\ref{Eq:OptimumU_for_gWwhite}), for which, maximal Dense coding capacity is achieved for given parameter values of $a$, $b$, and $s$. In this case, the maximum capacity becomes 
\begin{eqnarray}
    \mathcal{C}(\sigma_{AB}) = 2-H(\{\frac{s}{4},\frac{s}{4},a(1-s)+\frac{s}{4},1-\frac{3s}{4}-a(1-p)\})
    \label{Eq:Cap_whitenoise}
\end{eqnarray}
.$\blacksquare$\\
Suppose $J^*_x$ and $J^*_y$ are the optimal unitary parameter for which maximum amount of dense codeability can be activated in $\sigma_{AB}$.
Equating the eigenvalues of $\sigma_B$, we obtain,
\begin{eqnarray}
    && a(1-s)\cos{2(J^*_x-J^*_y)+(1-a-2b)(1-s)\cos{2(J^*_x+J^*_y)} = 0}\nonumber\\
    &\implies& (1-2b)\cos{2J^*_x}\cos{2J^*_y} - (1-2a-2b)\sin{2J^*_x}\sin{2J^*_y} = 0\nonumber\\
    &\implies& \tan{2 J^*_x} \tan{2 J^*_y} = \frac{1-2b}{1-2a-2b}
    \label{Eq:OptimumU_for_gWwhite}
\end{eqnarray}

\section{Calculations of part $2$ of Theorem $5$}\label{A3}
Suppose $J^*_x$ and $J^*_y$ are the optimal unitary parameter for which maximum amount of dense codeability can be activated in $\sigma_{AB}$. The eigenvalues for $\sigma_{AB}$ are found to be: $\{0,a-\frac{(a^2-(1-a)b)s}{a+b},\frac{1}{2}(1-a(1-s)-\frac{bs}{a+b}-k,\frac{1}{2}(1-a(1-s)-\frac{bs}{a+b}+k)\}$, where $k=\frac{\sqrt{(a+b-a^2(1-s)-ab(1-s)-bs)^2-4a(1-a-b)(1-s)(a+b)s}}{a+b})$. The eigenvalues for $\sigma_{B}$ are: $\Bigg\{\frac{1}{2}\Big(1-\frac{((a+b)(1-2b+2bs)-2as)\cos{2J_x}\cos{2J_y}}{a+b}+((1-2a-2b)-2s(1-a-b))\sin{2J_x}\sin{2J_y}\Big), \frac{1}{2}\Big(1+\frac{((a+b)(1-2b+2bs)-2as)\cos{2J_x}\cos{2J_y}}{a+b}-((1-2a-2b)-2s(1-a-b))\sin{2J_x}\sin{2J_y}\Big)\Bigg\}$.
Equating the eigenvalues of $\sigma_B$, we obtain,
\begin{eqnarray}
    && \frac{((a+b)(1-2b+2bs)-2as)\cos{2J_x}\cos{2J_y}}{a+b}-((1-2a-2b)-2s(1-a-b))\sin{2J_x}\sin{2J_y} = 0\nonumber\\
    &\implies& ((1-2b+2bs)-\frac{2as}{a+b})\cos{2J^*_x}\cos{2J^*_y} - ((1-2a-2b)-2s(1-a-b))\sin{2J^*_x}\sin{2J^*_y} = 0\nonumber\\
    &\implies& \tan{2 J^*_x} \tan{2 J^*_y} = \frac{(1-2b)+2s(b-\frac{a}{a+b})}{(1-2a-2b)-2s(1-a-b)}
\end{eqnarray}
\end{widetext}

\end{document}